\documentclass[10pt,letterpaper]{article}
\usepackage[top=0.85in,left=2.75in,footskip=0.75in,marginparwidth=2in, marginparsep=1.5cm]{geometry}

\usepackage[T2A,T1]{fontenc}
\usepackage[russian,english]{babel}
\usepackage{layouts}
\usepackage{nameref,hyperref}
\usepackage[bbgreekl]{mathbbol}
\usepackage{etoolbox}
\usepackage{mathtools}
\usepackage{csquotes}
\usepackage[right]{lineno}
\usepackage{lmodern}
\usepackage{tgheros}
\usepackage{threeparttable}
\usepackage{microtype}
\DisableLigatures[f]{encoding = *, family = * }
\usepackage{esint}
\usepackage{capt-of}
\usepackage[capitalise]{cleveref}
\usepackage{changepage}

\usepackage[aboveskip=1pt,labelfont=bf,labelsep=period,singlelinecheck=off]{caption}

\usepackage[backend=bibtex,
			natbib=true,
			style=numeric,
			sorting=none,
			maxbibnames=8,
			maxcitenames=2,
			giveninits=true,
			doi=true,
			url=false, isbn=false, eprint=false]{biblatex}
			
\makeatletter
\renewcommand{\@biblabel}[1]{\quad#1.}
\makeatother

\usepackage{lastpage,fancyhdr,graphicx}
\usepackage{epstopdf}
\fancyheadoffset[L]{2.25in}
\fancyfootoffset[L]{2.25in}

\usepackage{color}
\definecolor{Gray}{gray}{.25}

\usepackage{graphicx}
\usepackage{booktabs}
\usepackage[table]{xcolor}
\definecolor{Gray}{RGB}{100,100,100}
\definecolor{LightGray}{RGB}{245,245,249}
\definecolor{naturelight}{RGB}{248,231,215}
\definecolor{naturedark}{RGB}{204,139,136}

\definecolor{myblue}{RGB}{0,100,200}
\definecolor{myred}{RGB}{204,102,0}
\definecolor{mygreen}{RGB}{0,200,50}
\definecolor{cottoncandy}{RGB}{254,200,216}

\hypersetup{
	colorlinks,
	linkcolor=myred,
	citecolor=myblue,
	urlcolor=myblue
}

\usepackage{amsthm}
\usepackage{enumitem}

\newtheorem*{remark}{Remark}

\usepackage{amsmath}
\usepackage{siunitx}
\DeclareSIUnit\mmHg{mmHg}
\usepackage[scr=boondoxo]{mathalfa}

\newcommand{\mat}[1]{\boldsymbol{#1}}  
\newcommand{\vct}[1]{\boldsymbol{#1}}  
\newcommand{\T}{^{\text{T}}}
\newcommand{\inv}{^{-1}}
\newcommand*\dd{\mathop{}\!\mathrm{d}}

\newcommand{\NTC}{\omega^{N\bar{t}}}
\newcommand{\NHC}{\omega^{N\bar{h}}}
\newcommand{\oxygenv}{\omega^{n\bar{v}}}
\newcommand{\oxygenIF}{\omega^{n\bar{l}}}

\newcommand{\NP}{\mathsf{NP}}
\newcommand{\NPl}{\omega^{\mathsf{NP}\bar{l}}}
\newcommand{\NPv}{\omega^{\mathsf{NP}\bar{v}}}

\newcommand{\gammaNTdrug}{\gamma_{\mathsf{kill}}^{t}}
\newcommand{\gammaNHdrug}{\gamma_{\mathsf{kill}}^{h}}

\newcommand{\diffusivityNP}{D^{\NP l}}
\newcommand{\lymphaticfiltration}{\left(L_p \frac{S}{V}\right)^{\mathsf{ly}}}

\newcommand{\Asaltellimatrix}{\boldsymbol{A}}
\newcommand{\Bsaltellimatrix}{\boldsymbol{B}}
\newcommand{\ABisaltellimatrix}{\boldsymbol{A}_{\boldsymbol{B}}^{(i)}}
\newcommand{\ABjsaltellimatrix}{\boldsymbol{A}_{\boldsymbol{B}}^{(j)}}
\newcommand{\BAisaltellimatrix}{\boldsymbol{B}_{\boldsymbol{A}}^{(i)}}
\newcommand{\GPsurrogateprior}{f_{\text{GP}}}
\newcommand{\GPsurrogate}{f_{\text{GP},N}}
\newcommand{\GPmean}{m_{\textsf{GP},N}}
\newcommand{\GPx}{{\boldsymbol{X}}}
\newcommand{\GPXtrain}{\boldsymbol{\mathcal{X}}}
\newcommand{\GPXpredict}{\boldsymbol{\mathcal{X}}_*}
\newcommand{\GPYtrain}{\boldsymbol{\mathcal{Y}}}
\newcommand{\NGP}{{N_{\text{GP}}}}
\newcommand{\dmax}{d_{\textsf{max}}}
\newcommand{\Xvector}{\boldsymbol{X}}
\newcommand{\datatrain}{\mathcal{D}}
\newcommand{\Ex}[2]{{\mathbb{E}}_{#1}\left[#2\right]}
\newcommand{\Var}[2]{\mathbb{V}_{#1}\left[#2\right]}
\newcommand{\biomodel}{biomechanical model }
\newcommand{\biomodelend}{biomechanical model}
\newcommand{\biomodels}{biomechanical models }

\usepackage{setspace}
\usepackage{tabularx}
\usepackage{sansmath}
\usepackage{listings}
\usepackage{caption}

\DeclareCaptionFormat{mylst}{\hrule#1#2#3}
 
\usepackage{newfloat}
\DeclareFloatingEnvironment[fileext=frm,placement={tbp},name=Algorithm]{myfloat}
\newenvironment{myalgorithm}{\begin{myfloat}[tbp]}{\end{myfloat}}

\usepackage[font=sf]{caption}
\usepackage{nomencl}
\makenomenclature

\usepackage{etoolbox}
\renewcommand\nomgroup[1]{\item[\bfseries
  \ifstrequal{#1}{A}{Dimensions and corresponding indices}{\ifstrequal{#1}{C}{Sobol method related symbols}{\ifstrequal{#1}{B}{Random quantities}{\ifstrequal{#1}{D}{Metamodel related symbols}{\ifstrequal{#1}{O}{Other symbols}{}}}}}]}

\newcommand{\nomdimensions}[1][]{\nomenclature[A#1]}
\newcommand{\nomrandomqua}[1][]{\nomenclature[B#1]}
\newcommand{\nomsobol}[1][]{\nomenclature[C#1]}
\newcommand{\nommeta}[1][]{\nomenclature[D#1]}

\usepackage{csquotes}
\usepackage{pdfpages} % include pdfs
\usepackage{pgffor} % for loops

\makeatletter
\AtBeginDocument{\let\LS@rot\@undefined}
\makeatother

\def\supplementfilename{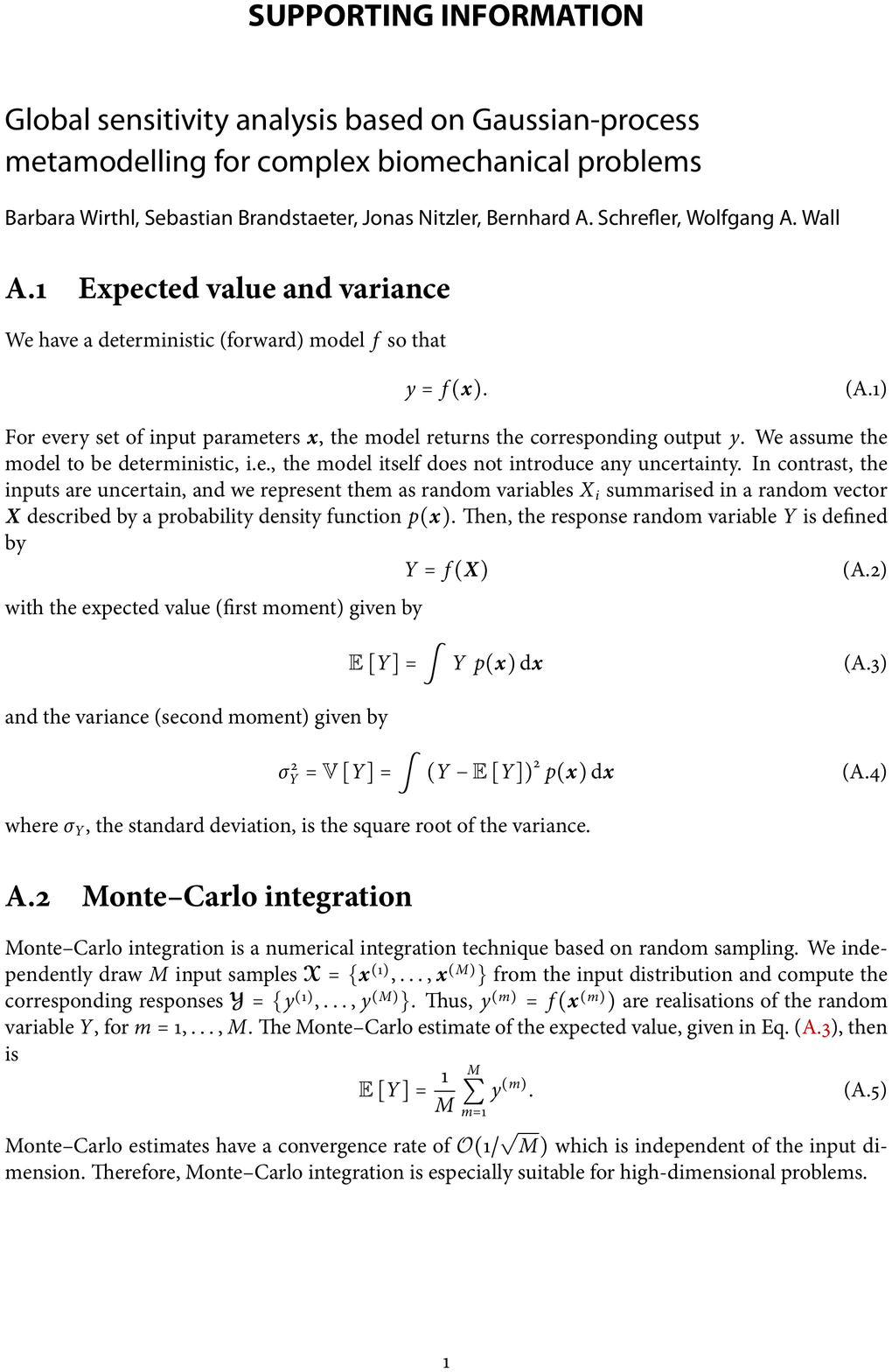}

\pdfximage{\supplementfilename}
\def\numbersupplementpages{\the\pdflastximagepages}

\begin{document}
\lstdefinestyle{interfaces}{
    float=bp,
    floatplacement=bp,
    abovecaptionskip=-5pt
}

\vspace*{0.35in}
{
\sffamily
\begin{flushleft}
    {\LARGE
        \setstretch{2}
        Global sensitivity analysis based on Gaussian-process metamodelling for complex biomechanical problems
    }
    \newline
    \\
    Barbara Wirthl\textsuperscript{1,*},
    Sebastian Brandstaeter\textsuperscript{1,2},
    Jonas Nitzler\textsuperscript{1,3},
    Bernhard A. Schrefler\textsuperscript{4,5},
    Wolfgang A. Wall\textsuperscript{1}
    \\
    \bigskip
    {
        \small
        \textbf{1} Institute for Computational Mechanics, Technical University of Munich, Garching b. München, Germany
        \\
        \textbf{2} Institute of Continuum and Materials Mechanics, Hamburg University of Technology, Hamburg, Germany
        \\
        \textbf{3} Professorship for Data-Driven Materials Modeling, Technical University of Munich, Garching b. München, Germany
        \\
        \textbf{4} Department of Civil, Environmental and Architectural Engineering, University of Padua, Padua, Italy
        \\
        \textbf{5} Institute of Advanced Study, Technical University of Munich, Garching b. München, Germany
        \\
        \bigskip
        * B. Wirthl, Email: barbara.wirthl@tum.de
    }

\end{flushleft}

{
\setlength{\parindent}{0cm}

{\Large \textbf{Abstract} \smallskip}

Biomechanical models often need to describe very complex systems, organs or diseases, and hence also include a large number of parameters.
One of the attractive features of physics-based models is that in those models (most) parameters have a clear physical meaning.
Nevertheless, the determination of these parameters is often very elaborate and costly and shows a large scatter within the population.
Hence, it is essential to identify the most important parameter (worth the effort) for a particular problem at hand.
In order to distinguish parameters which have a significant influence on a specific model output from non-influential parameters, we use sensitivity analysis, in particular the Sobol method as a global variance-based method.
However, the Sobol method requires a large number of model evaluations, which is prohibitive for computationally expensive models.
We therefore employ Gaussian processes as a metamodel for the underlying full model.
Metamodelling introduces further uncertainty, which we also quantify.
We demonstrate the approach by applying it to two different problems:
nanoparticle-mediated drug delivery in a complex, multiphase tumour-growth model, and arterial growth and remodelling.
Even relatively small numbers of evaluations of the full model suffice to identify the influential parameters in both cases and to separate them from non-influential parameters.
The approach also allows the quantification of higher-order interaction effects.
We thus show that a variance-based global sensitivity analysis is feasible for complex, computationally expensive biomechanical models.
Different aspects of sensitivity analysis are covered including a transparent declaration of the uncertainties involved in the estimation process.
Such a global sensitivity analysis not only helps to massively reduce costs for experimental determination of parameters but is also highly beneficial for inverse analysis of such complex models.
}
}

\newpage

\renewcommand{\nomname}{Key nomenclature}
\mbox{}

\nomrandomqua{\(\vct{x}  \in \mathbb{R}^D \)}{Deterministic vector of input parameters}
\nomrandomqua{\(y \in \mathbb{R}\)}{Deterministic scalar output of interest}
\nomrandomqua{\(X_i\)}{Uncertain input parameter (RV)}
\nomrandomqua{\(\vct{X}\)}{Vector of $D$ uncertain input parameters}
\nomrandomqua{\(Y\)}{Uncertain output of interest (RV)}
\nomrandomqua{\(\mathcal{X}\)}{Training samples}

\nomdimensions{\(D\)}{Number of uncertain input parameters, index $i$}
\nomdimensions{\(N\)}{Number of training samples}
\nomdimensions{\(M\)}{Number of Monte--Carlo samples, index $m$}
\nomdimensions{\(\NGP\)}{Number of metamodel realisations, index $k$}
\nomdimensions{\(B\)}{Number of bootstrap samples, index $b$}

\nomsobol{\(S^i\)}{First-order Sobol index for the $i$-th input parameter}
\nomsobol{\(S^{Ti}\)}{Total-order Sobol index for the $i$-th input parameter}
\nomsobol{\(S^{ij}\)}{Second-order Sobol index for input parameters $i$ and $j$}
\nomsobol{\(\hat{S}^\diamond\)}{Estimate of Sobol index of any order $\diamond$}

\nommeta{\(\GPsurrogate(\vct{\GPx})\)}{Trained Gaussian process metamodel}
\nommeta{\(\GPmean(\GPx)\)}{Predictive posterior mean of the Gaussian process with optimised hyperparameters trained on $N$ training samples}
\nommeta{\(k_N(\GPx, \GPx')\)}{Predictive posterior covariance kernel of the Gaussian process with optimised hyperparameters trained on $N$ training samples}

\printnomenclature

\newpage

\section{Introduction}

Over the past few decades, computational \biomodels have become an essential tool in research.
The goal of these models is to allow predictions so as to better understand the underlying biological system or to support decision-making in a medical context, e.g., to choose the most efficient therapy for a specific patient.
Nevertheless, the output of such models is inherently subject to uncertainty for various reasons:
first, the underlying biological process is stochastic---which is particularly true for oncophysics and cancer treatment models (for example, branching process models for cancer \cite{Durrett2013, Durrett2015}, or stochastic models for immunotherapy of cancer \cite{Baar2016}).
Second, the experimental data used to calibrate models is uncertain \cite{White2004}.
Third, the computational model itself includes sources of uncertainty, including the assumptions made to set up the model, other simplifications, or the input parameters \cite{Saltelli2019}.

When analysing the uncertainty of the model output, we distinguish uncertainty analysis from sensitivity analysis \cite{Saltelli2019}:
uncertainty analysis quantifies the uncertainty in the model output by propagating input uncertainties, via the model, onto the output \cite[p.~262]{Saltelli2008}.
Sensitivity analysis, on the other hand, apportions the uncertainty in the model output to different sources of uncertainty in the model input \cite[p.~45]{Saltelli2004}.
Inputs of interest can generally include not only model parameters but also boundary and initial conditions, assumptions, and constraints \cite{Razavi2021}.
In the context of sensitivity analysis, those inputs of interest are commonly referred to as \textit{factors}.
Here, we only consider model parameters as sources of uncertainty and refer to those as \textit{input parameters}.

In this study, the goal is threefold:
\begin{itemize}[nosep]
    \item Identify the most influential parameters on which further experimental estimation should focus (called factor prioritisation or ranking).
    \item Identify parameters with little or no effect, which can thus be set to fixed values within their range (called factor fixing or screening).
    \item Identify and quantify the interaction between parameters.
\end{itemize}
This knowledge expedites the efficient design of future computational and experimental studies, while avoiding wasting resources on determining non-influential parameters.

We propose to apply a special type of sensitivity analysis to achieve the goals just described.
One way of quantifying the sensitivity of the model output $Y$ on the input parameter $X_i$ is to calculate the partial derivative $\partial Y/ \partial X_i$.
In practice, this involves choosing a base point $\vct{X}^*$ and then perturbing one factor at a time while keeping all remaining factors fixed.
This results in a local sensitivity measure at the base point $\vct{X}^*$ which only explores one point of the input space and thus results in a deficient sensitivity analysis \cite{Saltelli2019}.
In contrast, the Elementary Effects method (also called Morris method \cite{Morris1991}) is not limited to one single point but explores the whole input space.
It thereby overcomes the major limitation of local methods, while only requiring a relatively small number of model evaluations.
While the Elementary Effects method is a global sensitivity analysis method, it only provides semi-quantitative information and is typically used for factor fixing \cite{Saltelli2004,Saltelli2008}.
However, it cannot detect and quantify interactions between parameters and nonlinearities \cite{Qian2020}.
In this work, we focus on complex biomechanical problems in which interactions between the different parameters can be expected.
We therefore need a global method that can provide more detailed information.

\begin{figure}[tbp]
    \includegraphics[width=\textwidth]{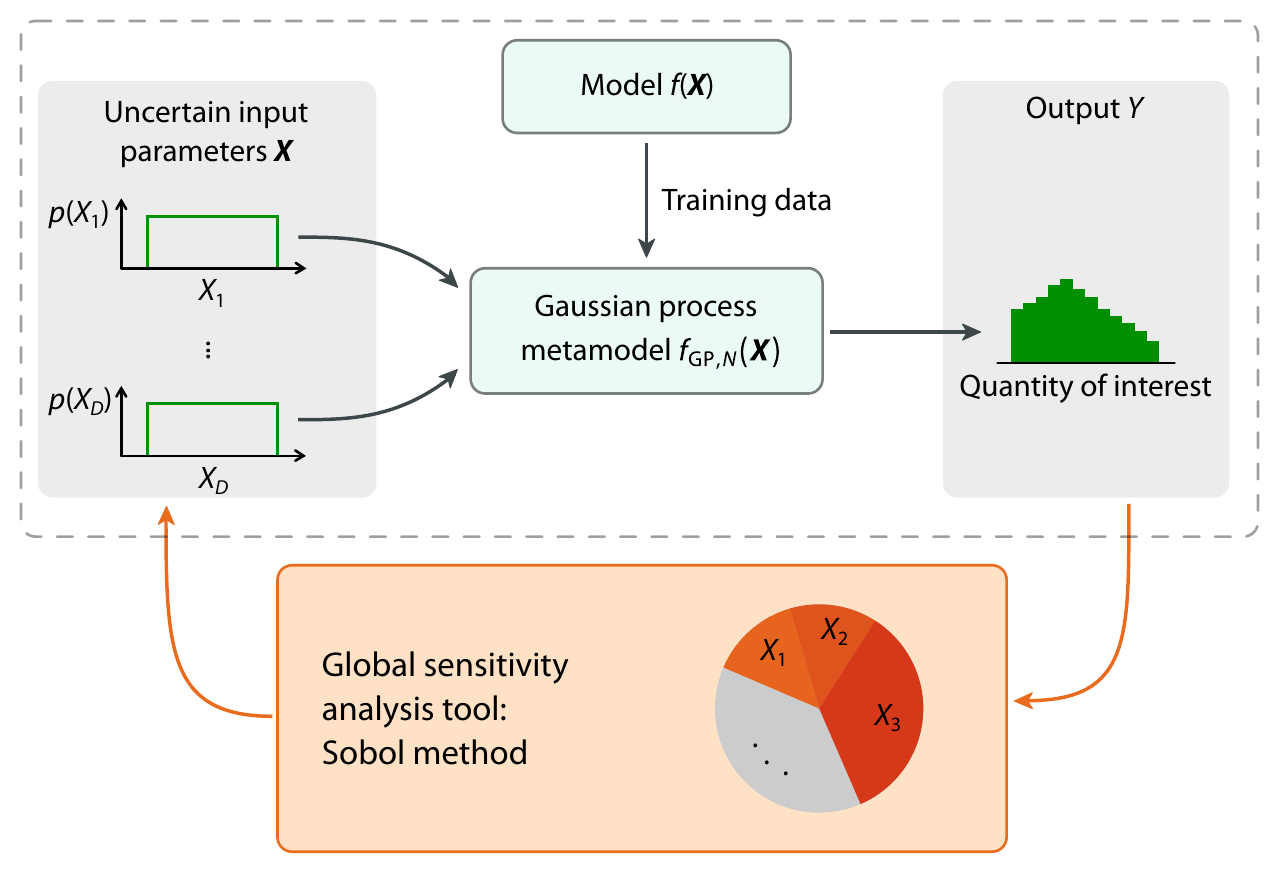}
    \caption{\color{Gray} \sffamily \textbf{Overview of sensitivity analysis with a Gaussian process metamodel}.
        The Gaussian-process metamodel $\GPsurrogate(\vct{X})$ is trained on a data set generated by the full model $f(\vct{X})$.
        Uncertain input parameters $X_i$ are propagated via the metamodel and result in an uncertain output $Y$.
        The Sobol method, as an example of a global sensitivity analysis tool, decomposes the output variance into portions attributed to the input parameters.
    }
    \label{fig:Overview}
\end{figure}

Our method of choice is the Sobol\footnotemark method \cite{Sobol1993, Sobol2001}, which is a variance-based global sensitivity analysis method that decomposes the output variance into portions attributed to the input parameters (see \cref{fig:Overview}).
The downside is that it requires many model evaluations, which quickly becomes computationally prohibitive in the case of complex models.
We propose to introduce Gaussian processes \cite{Rasmussen2006} as a metamodel for the full model to mitigate the problem of computationally expensive model evaluations.
Since the use of a metamodel introduces a further source of uncertainty in the sensitivity analysis, we estimate the uncertainty following the approach presented by Le Gratiet \textit{et al.}~\cite{LeGratiet2014}.
After the full \biomodel is substituted by the metamodel, we can calculate the Sobol indices based on Monte--Carlo integration.
The uncertainty related to metamodelling and the uncertainty related to Monte--Carlo integration are analysed both separately and in total.

\footnotetext{The Sobol method was proposed by and is named after Ilya Meyerovich Sobol' (Russian: \foreignlanguage{russian}{Илья Меерович Соболь}) to whose last name an apostrophe is appended in English to transliterate the Russian letter \foreignlanguage{russian}{ь}.
    To avoid confusion with the apostrophe used in English grammar, we omit the apostrophe when referring to the Sobol method.
}

So far, the approach suggested by \cite{LeGratiet2014} has been applied to different computational models:
an individual-based model simulation of microbial communities \cite{Oyebamiji2017}, a mathematical model of renal fibrosis \cite{Karagiannis2020}, climate change simulations \cite{Ross2020}, and numerical wind-turbine models \cite{Hirvoas2021}.
Moreover, \cite{SahliCostabal2019} applied the same idea to the calculation of Elementary Effects of a heart model.
In \cite{Oyebamiji2017, Karagiannis2020, SahliCostabal2019, Ross2020}, the authors state that they used the method, but no analysis of the associated uncertainties was presented.
Only \cite{Hirvoas2021} quantified the uncertainty related to the metamodel and the uncertainty related to Monte--Carlo integration separately.

Our goal is to present the complete workflow of estimating Sobol indices based on Gaussian processes as a metamodel including the uncertainties:
we demonstrate how to apply the approach suggested by \cite{LeGratiet2014} to two different biomechanical models, and we also assess the performance of the method when applied to such complex examples.

The article is structured as follows:
we first introduce the reader to the Sobol method and to Gaussian processes in general.
The Gaussian-process metamodel is then used to estimate the Sobol indices, including separate estimates for uncertainty related to Monte--Carlo integration and uncertainty related to the metamodel based on \cite{LeGratiet2014}.
As an example, we subsequently demonstrate how to apply this approach to a multiphase model of nanoparticle-mediated drug delivery in a solid tumour \cite{Kremheller2018,Kremheller2019,Wirthl2020} and assess its performance in detail.
Finally, we conclude with an outlook on a different complex \biomodelend, i.e., a homogenised constrained mixture model of arterial growth and remodelling \cite{Cyron2016,Cyron2017,Braeu2019}.

\section{Global sensitivity analysis based on Gaussian-process metamodelling}

\subsection{The Sobol method}

In its most general form, a model $f$ is a functional representation of the relevant physical process.
The model calculates an output $y = f(\vct{x}) \in \mathbb{R}$ for any given realisation $\vct{x} \in \mathbb{R}^D$  of the uncertain input parameters $\vct{X}$, with $D$ being the number of parameters (see upper left side of \cref{fig:Overview}).
We assume that the random vector $\vct{X}$ summarises the input parameters $X_1, \dots, X_D$, which are independent random variables.
The probability distribution of $\vct{X}$ is described by the probability density function $p(\vct{x})$.
In the following, we assume that the input parameters are uniformly distributed, with no loss of generality.
Common alternatives are the normal distribution or the log-normal distribution, among many others.

Our goal is to investigate the sensitivity of the model output to the uncertain input parameters $\vct{X}$.
Because of the randomness in the input parameters, the model output $Y$ is also a random variable defined as
\begin{equation}
    Y = f(\vct{X}) = f(X_1, X_2, \dots, X_D)
\end{equation}
(see upper right side of \cref{fig:Overview}).
The output distribution can (partially) be described by its first two moments:
the expected value $\Ex{}{Y}$, and the variance $\sigma^2_Y = \Var{}{Y}$ (with $\Ex{}{\cdot}$ denoting the expectation operator, $\Var{}{\cdot}$ the variance operator, and $\sigma_Y$ the standard deviation as given in Supplement A.1).

One way of characterising sensitivity is to decompose the variance of the output $\Var{}{Y}$ into portions ascribed to the individual input parameters.
A common sensitivity analysis method based on the decomposition of variance is the Sobol method \cite{Sobol1990}.
The core idea is to decompose the output variance $\Var{}{Y}$ as
\begin{equation}
    \Var{}{Y}
    =
    \sum\limits_{i=1}^{D} V_i + \sum\limits_{i=1}^{D-1}\sum\limits_{j>i}^{D} V_{ij} + \dots + V_{12\dots D}
    \label{Eq:VarianceDecomposition}
\end{equation}
with the conditional variances given by
\begin{align*}
    V_i     & = \Var{X_i}{\Ex{\Xvector_{\sim i}}{Y | X_i}}                                                                    \\
    V_{ij}  & = \Var{X_i,X_j}{\Ex{\Xvector_{\sim i,j}}{Y | X_i , X_j}} - V_i - V_j                                            \\
    V_{ijk} & = \Var{X_i,X_j,X_k}{\Ex{\Xvector_{\sim i,j,k}}{Y | X_i, X_j, X_k}} - V_{ij} - V_{ik} - V_{jk} - V_i - V_j - V_k \\
            & \dots
\end{align*}
where $\Xvector_{\sim i}$ denotes the vector of all input parameters except $X_i$.
Thus, the output variance $\Var{}{Y}$ is the sum of variances contributed by input parameter $X_i$, including interactions with other parameters.
The idea now is to attribute the total variance to the individual input parameters according to their variance contribution.
Note that $V_{ij}$ is the variance contributed by the input parameters $X_i$ and $X_j$ but not expressed in $V_i$ nor $V_j$.
This is called the interaction of parameters $X_i$ and $X_j$.
Note that this decomposition assumes statistical independence of the input parameters $X_i$.

Because $\Var{X_i}{\Ex{\Xvector_{\sim i}}{Y | X_i}}$ is the portion of the output variance ascribed to input parameter $X_i$, we define the first-order Sobol index $S^i$ as
\begin{equation}
    S^i = \frac{\Var{X_i}{\Ex{\Xvector_{\sim i}}{Y | X_i}}}{\Var{}{Y}}.
    \label{Eq:FirstOrderSobolIndex}
\end{equation}
The numerator $\Var{X_i}{\Ex{\Xvector_{\sim i}}{Y | X_i}}$ describes the extent to which the output variance $\Var{}{Y}$ would be reduced if the parameter $X_i$ was fixed.
A parameter $X_i$ with a high first-order index $S^i$ should, hence, have priority when determining parameters based on experiments so as to efficiently reduce the overall uncertainty of the model.
The first-order Sobol index is typically used to identify the most influential parameters, which is our first goal \cite{Qian2020}.
Moreover, a parameter with a high first-order index $S^i$ is more likely to be identifiable from experiments, but can still be non-identifiable \cite{Razavi2021}:
to decide whether a parameter is identifiable or not, an identifiability analysis is required, which complements the sensitivity analysis (see \cite{Guillaume2019} for an overview of identifiability analysis).

The question now arises as to whether $S^i = 0$ is also sufficient to conclude that a parameter has no influence.
In fact, this is not the case because the parameter might be involved in interactions with other parameters.
A parameter may have no effect if it is varied alone;
however, this may be different when it is varied in combination with another parameter, or even with several other parameters.
An additional sensitivity measure that includes higher-order interaction effects is needed to identify non-influential parameters.
The total-order Sobol index is, therefore, defined as
\begin{equation}
    S^{Ti} = \frac{\Ex{\Xvector_{\sim i}}{\Var{X_i}{Y | \Xvector_{\sim i}}}}{\Var{}{Y}}
    =
    1 -
    \frac{\Var{\Xvector_{\sim i}}{\Ex{X_i}{Y | \Xvector_{\sim i}}}}{\Var{}{Y}}
    .
    \label{Eq:TotalSobolIndex}
\end{equation}
In this case, the numerator $\Ex{\Xvector_{\sim i}}{\Var{X_i}{Y | \Xvector_{\sim i}}}$ describes the expected output variance that would be left if all parameters but $X_i$ were to be determined \cite{Qian2020}.
If---and only if---this expected output variance is close to zero, is the parameter $X_i$ non-influential.
The total-order index describes the total contribution of the parameter $X_i$ to the output $Y$:
this includes the first-order effect plus any higher-order effects that arise from interactions.
The difference $S^{Ti} - S^i$, then, indicates interaction effects between factor $X_i$ and any other factor \cite[p.~167]{Saltelli2008}.
As mentioned above, the total-order index is particularly helpful in the context of factor fixing:
if $S^{Ti} = 0$ (or is in practice sufficiently small), the parameter $X_i$ is non-influential and can be fixed anywhere in its input range without affecting the output variance.

So, the first-order and the total-order Sobol indices serve our first two goals: identify the most influential and the non-influential parameters.
To additionally identify interactions between two specific parameters $X_i$ and $X_j$---which is our third goal---we define the second-order Sobol index as
\begin{equation}
    S^{ij} = \frac{\Var{X_i,X_j}{\Ex{\Xvector_{\sim i,j}}{Y | X_i , X_j}}}{\Var{}{Y}}
    - S^i - S^j.
    \label{Eq:SecondOrderSobolIndex}
\end{equation}

Finally, dividing \cref{Eq:VarianceDecomposition} by $\Var{}{Y}$ and inserting \cref{Eq:FirstOrderSobolIndex,,Eq:SecondOrderSobolIndex} leads to
\begin{equation}
    \sum\limits_{i=1}^{D} S^i + \sum\limits_{i=1}^{D-1}\sum\limits_{j>i}^{D} S^{ij} + \ldots + S^{12\dots D} = 1.
    \label{Eq:SumIndices}
\end{equation}
All sensitivity indices, thus, sum up to 1;
furthermore, they are non-negative.
This leads to an interesting implication which is worth noting:
even when we have a large number of parameters, we cannot have a large number of influential parameters.
If all $D$ parameters are equally influential, each can only contribute $1/D$ of the variance.
If, however, a few parameters have a strong influence on the output $Y$, the remaining parameters can contribute even less (see~\cite[Sec. 2.4.3]{Saltelli2008}).
As \cite{Box1986} stated:
only a small subset of parameters significantly influences one specific system output (\enquote{sparsity of factors} principle).
It should also be noted that the total-order indices $S^{Ti}$ do not, in general, sum up to 1.

\subsection{Numerical approximation of Sobol indices}

To estimate the Sobol indices according to \cref{Eq:FirstOrderSobolIndex,,Eq:TotalSobolIndex,,Eq:SecondOrderSobolIndex}, we need to compute conditional variances, e.g., $\Var{X_i}{\Ex{\Xvector_{\sim i}}{Y | X_i}}$, which involves evaluating multidimensional integrals in the space of the input parameters $\mathbb{R}^D$.
Numerical integration based on quadrature rules becomes prohibitively expensive as the number of input space dimensions increases.
This is why Monte--Carlo integration is employed, the accuracy of which is independent of the number of input space dimensions \cite{Asmussen2007}.
For each single integral, Monte--Carlo integration involves evaluating $M$ Monte--Carlo samples:
to compute, for example, $\Var{X_i}{\Ex{\Xvector_{\sim i}}{Y | X_i}}$ one would need $M$ samples to calculate the inner expectation and then repeat this $M$ times to calculate the outer variance, resulting in a computational cost of $\mathcal{O}(M^2)$ \cite[p. 164]{Saltelli2008}.
Since $M$ usually has to be large\footnotemark, this is impractical, especially considering that we would need to evaluate the full model $f$ for each Monte--Carlo sample.
To make the estimation of Sobol indices more efficient, \cite{Ishigami1990} rewrote the multidimensional integral so that it can be computed using a single Monte--Carlo loop (summarised in Supplement A.2).

\footnotetext{The error of the Monte--Carlo estimate for the expectation is proportional to $\frac{\sqrt{\Var{}{g}}}{\sqrt{M}}$, with $g$ denoting the integrand.
    If we assume $\Var{}{g}$ to be fixed, we have to increase the number of Monte--Carlo samples $M$, and the error of the estimate thus decreases by $\frac{1}{\sqrt{M}}$ \cite{Jarosz2008}.
}

To make the best use of the model evaluations, we employ the efficient algorithms suggested by \cite{Saltelli2002}:
to estimate the first and the total-order indices, we generate $M$ samples row-wise concatenated as a matrix $\Asaltellimatrix$ and further $M$ samples concatenated as a matrix $\Bsaltellimatrix$.
This results in two independent $M \times D$ matrices.
We introduce a third matrix $\ABisaltellimatrix$ for each input space dimension $i = 1, \dots, D$, where all columns are taken from $\Asaltellimatrix$ except the $i$-th column, which is taken from $\Bsaltellimatrix$.
One sample $(\Bsaltellimatrix)_m$ and the corresponding sample $(\ABisaltellimatrix)_m$ have $X_{i}$ in common but differ in all other parameters $\vct{X}_{\sim i}$.

To calculate the first-order index, we then use the estimator proposed by Saltelli \textit{et al.}~\cite{Saltelli2010}
\begin{equation}
    \Var{X_i}{\Ex{\Xvector_{\sim i}}{Y | X_i}}
    \approx
    \frac{1}{M} \sum\limits_{m=1}^{M}
    f(\Bsaltellimatrix)_m \left(f(\ABisaltellimatrix)_m - f(\Asaltellimatrix)_m \right),
    \label{Eq:Saltelli2010}
\end{equation}
and for the total-order index, we use the estimator proposed by Jansen~\cite{Jansen1999}:
\begin{equation}\label{Eq:Jansen1999}
    \Ex{\Xvector_{\sim i}}{\Var{X_i}{Y | \Xvector_{\sim i}}}
    \approx
    \frac{1}{2M} \sum\limits_{m=1}^{M}
    \left(f(\Asaltellimatrix)_m - f(\ABisaltellimatrix)_m \right)^2.
\end{equation}
Alternative forms were presented in \cite{Sobol1993, Jansen1999, Sobol2007, Janon2014a}, among others.
The denominator is estimated as
\begin{equation}
    \Var{}{Y}
    \approx
    \mathcal{V}[f([\Asaltellimatrix\; \Bsaltellimatrix])],
\end{equation}
where we estimate the variance of the output as the sample variance $\mathcal{V}$ of evaluations of all samples $\Asaltellimatrix$ and $\Bsaltellimatrix$.
This yields better results, i.e., an estimator with lower variance, compared to $\Var{}{Y} \approx \mathcal{V}[f(\Asaltellimatrix)]$ alone \cite{Saltelli2002}.

In addition to the first and total-order indices, we estimate the second-order indices as proposed by Saltelli~\cite{Saltelli2002}:
\begin{equation}
    \Var{X_i,X_j}{\Ex{\Xvector_{\sim i,j}}{Y | X_i , X_j}}
    \approx
    \frac{1}{M} \sum\limits_{m=1}^{M} f(\BAisaltellimatrix)_m \; f(\ABjsaltellimatrix)_m - f(\Asaltellimatrix)_m \; f(\Bsaltellimatrix)_m
    \label{Eq:SecondOrderEstimator}
\end{equation}
where the matrix $\BAisaltellimatrix$ is built similar to $\ABisaltellimatrix$.
More details on different sensitivity-index estimators can be found in \cite{Saltelli2002, Saltelli2010}, among others.

We hence have to evaluate our model $f$ at all samples of the triplet $\Asaltellimatrix$, $\Bsaltellimatrix$ and $\ABisaltellimatrix$ (and additionally $\BAisaltellimatrix$ if second-order indices are included).
This means $2M$ simulations are needed for computing $f(\Asaltellimatrix)$ and $f(\Bsaltellimatrix)$ plus $D \cdot M$ simulations needed for computing $f(\ABisaltellimatrix)$ for $i = 1, \dots, D$.
The cost of first and total-order indices is, hence, $M (D + 2)$ simulations.
If second-order indices are included, we need an additional $D \cdot M$ simulations for $f(\BAisaltellimatrix)$, resulting in $M (2D + 2)$ simulations in total (for more details see Supplement A.3 and the original publication by \cite{Saltelli2002}).
In practice, Quasi--Monte--Carlo (QMC) integration is often used to generate the samples because of its superior rate of convergence compared to Monte--Carlo integration \cite{Niederreiter1992}.

\subsection{Gaussian process metamodels}

As just described, Monte--Carlo integration to estimate the Sobol indices requires a large number of sample evaluations and is thus computationally prohibitive if the evaluation of the underlying model is expensive.
We therefore use a metamodel (also known as surrogate model or emulator) as an approximation of the full model.
Classically used metamodels include polynomials, splines, neural networks, polynomial chaos expansion, support vector regression, and Gaussian processes (GPs), among others \cite{Iooss2010,Cheng2020}.
Before a metamodel can be used for a sensitivity analysis, for example, it has to be trained to later ensure that it is a good approximation of the full model.

\begin{figure}[ptb]
    \includegraphics[width=\textwidth]{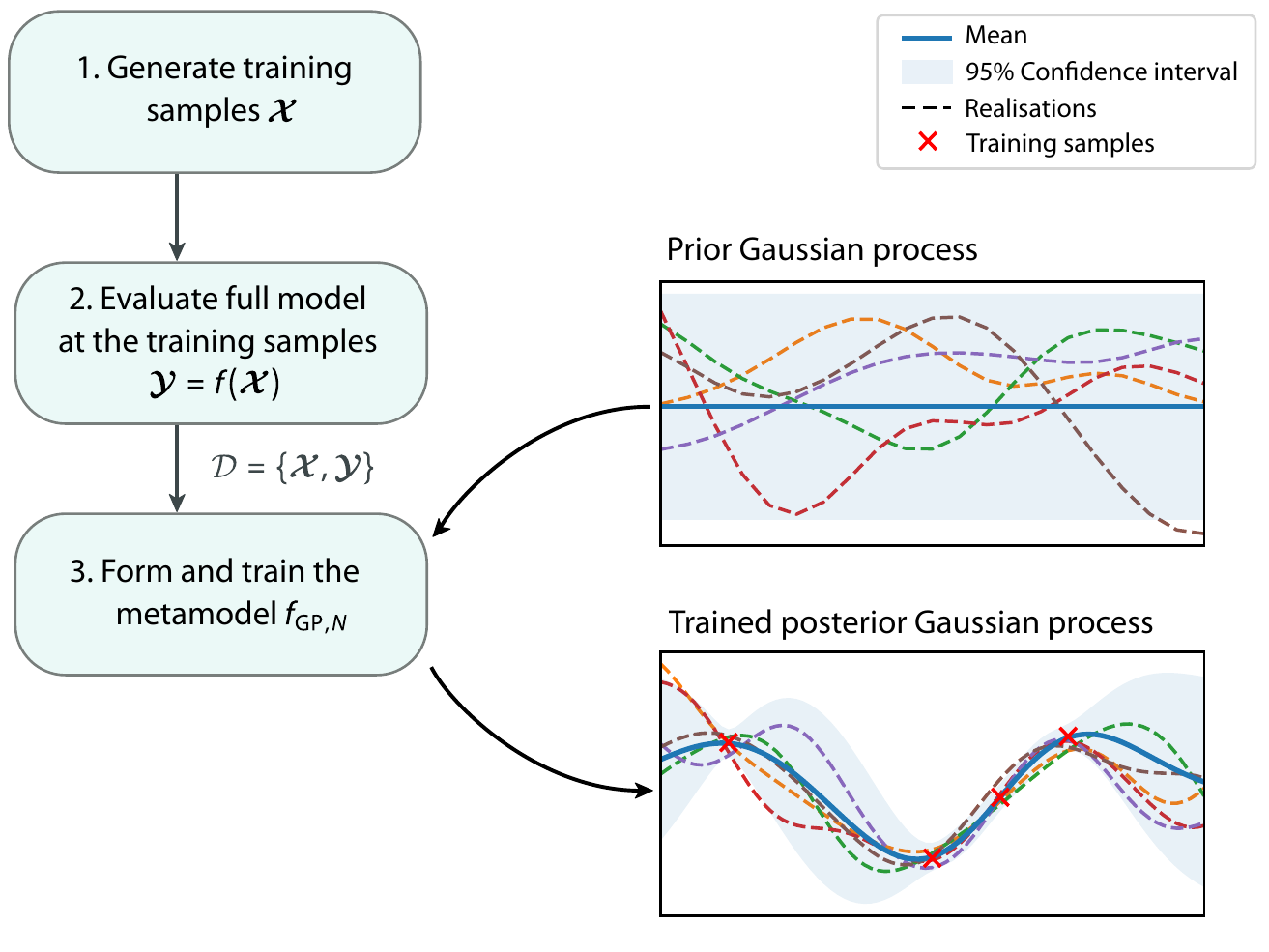}
    \caption{\color{Gray} \sffamily \textbf{Schematic overview of the steps involved in training the Gaussian process metamodel}.
        Conditioning the prior Gaussian process on the training samples results in the posterior Gaussian process.
    }
    \label{fig:OverviewPriorPosteriorGP}
\end{figure}

This process consists of three steps, which we first summarise (see \cref{fig:OverviewPriorPosteriorGP}) and, then, explain in more detail below:
\begin{enumerate}
    \item Generate $N$ training samples summarised in $\GPXtrain$ (resulting in an $N \times D$ matrix).
    \item Evaluate the full model at the training samples to obtain the corresponding response: $\GPYtrain = f(\GPXtrain)$ (resulting in an $N \times 1$ vector).
    \item Form and train the metamodel.
\end{enumerate}

First, we generate $N$ training samples that are summarised in the matrix $\GPXtrain$ ($N \times D$ matrix).
The choice of training samples has to provide a good coverage of the input space to later ensure a good predictive quality of the metamodel.
To this end, we use a QMC approach based on Sobol sequences \cite{Sobol1967} to generate the training samples.
\begin{remark}[Sequential design]
    A commonly used alternative to a QMC approach is Latin Hypercube Sampling (LHS) \cite{McKay1979}.
    \cite{Iooss2010} state that optimised LHS is particularly well-suited for metamodel fitting.
    However, sequential design is also important in the context of metamodelling:
    if the original number of training samples is not sufficient to achieve a good predictive quality of the metamodel, additional samples can be added while still making use of the original training samples.
    Since only advanced LHS methods \cite{Xiong2009, Zhou2019a} enable sequentially adding new points, while this is straightforward with QMC schemes \cite{Kucherenko2015}, we use a QMC approach.
\end{remark}

Subsequently, we evaluate the full model at each training sample: $\GPYtrain = f(\GPXtrain)$ with $\GPYtrain$, hence, being an $N \times 1$ vector.
This results in the training data set
\begin{equation}
    \mathcal{D} = \left\lbrace \GPXtrain, \GPYtrain \right\rbrace
    \label{Eq:dataset}
\end{equation}
where each row corresponds to one training point.

Finally, we have to \textit{train} the metamodel.
We use a GP $\GPsurrogateprior$ as metamodel similar to \cite{Oakley2004,Marrel2009,LeGratiet2014} and summarise what training means for GPs below.
Note that we only include a compact overview of the most relevant concepts used in this paper.
For more details, the reader is referred e.g. to \cite{Rasmussen2006}.

A GP defines a distribution over functions such that any finite set of function values $\GPsurrogateprior(\vct{x}^{(1)})$, $\GPsurrogateprior(\vct{x}^{(2)})$, \dots, $\GPsurrogateprior(\vct{x}^{(n)})$ has a joint Gaussian distribution \cite{Rasmussen2006, Duvenaud2014}.
From a Bayesian point of view, we distinguish between prior and posterior:
the prior GP reflects our beliefs about the metamodel before seeing any (training) data, and the posterior GP is then conditioned on the (training) data, i.e. includes the knowledge from the data (see \cref{fig:OverviewPriorPosteriorGP}).
This conditioning on the data is what we refer to as training.

The prior GP $\GPsurrogateprior(\vct{X})$ is given by
\begin{equation}
    \label{Eq:GaussianProcessPrior}
    \GPsurrogateprior(\vct{\GPx}) \sim \text{GP}(m_{\text{GP}}(\GPx), k(\GPx, \GPx'))
\end{equation}
and is completely specified by its mean function $m_{\text{GP}}(\GPx)$ and its covariance function $k(\GPx, \GPx')$ between all possible pairs $(\GPx, \GPx')$.
The covariance function is a positive definite kernel, e.g., the squared exponential covariance function (also called radial basis function)
\begin{equation}
    \label{Eq:RBF}
    k(\GPx, \GPx')
    =
    \sigma_f^2 \exp\left( -\frac{1}{2\ell^2} || \GPx - \GPx' ||^2 \right)
\end{equation}
with the characteristic length scale $\ell$, variance parameter $\sigma_f$ and $|| \cdot ||$ denoting the Euclidean L2-norm.
We assume that the prior mean function is zero: $m_{\text{GP}}(\GPx) = \vct{0}$, which is common practice and does not limit the GP model, as any uncertainty about the mean function can be included in the choice of a covariance function \cite{Duvenaud2014}.
Different covariance functions exist and can be combined, for example, through multiplication or addition.
\cite[Chap. 2]{Duvenaud2014} presents a concise overview of different covariance functions for GPs and how to use them to express the structure of the data.
The choice of a suitable covariance function is essential since the more a-priori knowledge goes into choosing the covariance function, the fewer data we need to train the metamodel \cite{Goodman2011}.

As described above, we observe the output at $N$ training points.
The goal, then, is to predict the output at $N_*$ new points summarised in $\GPXpredict$;
in our case, those new points (where we predict the output) will be the Monte--Carlo samples for the estimation of the Sobol indices.

Remember that \cref{Eq:GaussianProcessPrior} is only the prior distribution and does not yet incorporate our knowledge from the training data.
To obtain the posterior, we now condition the prior GP on our set of $N$ training data points.
This conditioning results in the key predictive equations
\begin{align}
    \GPsurrogate(\vct{\GPXpredict}) & \sim \text{GP}(m_{\text{GP},N}(\GPXpredict), k_N(\GPXpredict, \GPXpredict)) \text{ with} \\
    m_{\text{GP},N}(\GPXpredict)    & = \mat{K}_{*}\T \mat{K}_\varepsilon\inv \GPYtrain, \label{Eq:PredictiveMean}             \\
    k_N(\GPXpredict, \GPXpredict)   & =  \mat{K}_{**} - \mat{K}_{*}\T \mat{K}_\varepsilon\inv \mat{K}_{*}.
    \label{Eq:PredictiveVariance}
\end{align}
In \cref{Eq:PredictiveMean,,Eq:PredictiveVariance}, $\mat{K}_\varepsilon = \mat{K} + \sigma_y^2 \mat{I}$, where $\mat{K} = k(\GPXtrain,\GPXtrain)$ denotes the $N \times N$ matrix we obtain when evaluating the covariance function (given in \cref{Eq:RBF}) for all pairs of training points, similarly for $\mat{K}_{*} = k(\GPXtrain,\GPXpredict)$ and $\mat{K}_{**} = k(\GPXpredict, \GPXpredict)$.
We use $\sigma_y^2$ as an artificially introduced variable nugget term to alleviate numerical problems \cite{Pepelyshev2010, Andrianakis2012}.
The hyperparameters $\vct{\theta} = (\sigma_f, \ell)$ are optimised by maximising the log marginal likelihood using a gradient-based optimiser.
The log marginal likelihood is given by
\begin{equation}
    \label{Eq:LogMarginalLikelihood}
    \log p(\GPYtrain | \GPXtrain)
    =
    - \frac{1}{2} \GPYtrain\T \mat{K}_\varepsilon\inv \GPYtrain
    - \frac{1}{2} \log |\mat{K}_\varepsilon|
    - \frac{N}{2} \log 2 \pi
\end{equation}
with $|\cdot|$ denoting the determinant.
Maximising the log marginal likelihood given by \cref{Eq:LogMarginalLikelihood} with respect to the hyperparameters $\vct{\theta}$ automatically incorporates a trade-off between model fit and model complexity:
the first term in \cref{Eq:LogMarginalLikelihood} penalises the model's failure to describe the data while the second term penalises high model complexity.
Thus, this favours the least complex model that is able to explain the data \cite{Rasmussen2006}.

One advantage of employing GPs as a metamodel is that predictions can be computed exactly in a closed form \cite{Duvenaud2014} and that GPs inherently provide uncertainty measures over the predictions.
Moreover, one can incorporate a wide range of modelling assumptions into the choice of the covariance function.
However, note that computing the inverse in the first term in \cref{Eq:LogMarginalLikelihood} (and the determinant in the second term) is computationally expensive, i.e., on the order $\mathcal{O}(N^3)$.
This cubic complexity results in slow inference as the number of training samples increases.
One further challenge of using GPs as a metamodel is that they are susceptible to the curse of dimensionality:
as the dimensionality of the input space increases, the number of training samples required to train the metamodel grows exponentially \cite{Bengio2006, Tripathy2016} and the optimisation of hyperparameters $\vct{\theta}$ becomes impractical.

\begin{remark}[Advanced GPs]
    In case of large numbers of training samples and/or input space dimensions, various advanced GP metamodels are available:
    \cite{Liu2020} review approaches to improve the scalability of GPs to large data sets, e.g., by using stochastic variational inference \cite{Hensman2013};
    \cite{Tripathy2016} present an approach with built-in dimensionality reduction.
\end{remark}

\subsection{Estimation of Sobol indices and their uncertainty}
To estimate the Sobol indices, we now use the estimators given by \cref{Eq:Saltelli2010,,Eq:Jansen1999,,Eq:SecondOrderEstimator} and substitute the realisations of the full model $f$ with those of the trained GP metamodel $\GPsurrogate$ as suggested by \cite{LeGratiet2014}:
\begin{equation}
    \hat{S}^i = \frac{
        \frac{1}{M} \sum_{m=1}^{M}
        \GPsurrogate(\Bsaltellimatrix)_m \;
        (
        \GPsurrogate(\ABisaltellimatrix)_m - \GPsurrogate(\Asaltellimatrix)_m
        )
    }{
        \mathcal{V}[\GPsurrogate([\Asaltellimatrix\;  \Bsaltellimatrix])]
    },
    \label{Eq:SobolEstimatorFirstGP}
\end{equation}
\begin{equation}
    \hat{S}^{Ti} = \frac{
    \frac{1}{2M} \sum_{m=1}^{M}
    (\GPsurrogate(\Asaltellimatrix)_m - \GPsurrogate(\ABisaltellimatrix)_m )^2
    }{
    \mathcal{V}[\GPsurrogate([\Asaltellimatrix\;  \Bsaltellimatrix])]
    },
    \label{Eq:SobolEstimatorTotalGP}
\end{equation}
\begin{equation}
    \hat{S}^{ij} = \frac{
        \frac{1}{M} \sum_{m=1}^{M}
        \GPsurrogate(\BAisaltellimatrix)_m \; \GPsurrogate(\ABjsaltellimatrix)_m
        -
        \GPsurrogate(\Asaltellimatrix)_m \; \GPsurrogate(\Bsaltellimatrix)_m
    }{
        \mathcal{V}[\GPsurrogate([\Asaltellimatrix\;  \Bsaltellimatrix])]
    }
    -
    \hat{S}^i
    -
    \hat{S}^j
    \label{Eq:SobolEstimatorSecondGP}
\end{equation}
with $M$ again being the number of Monte--Carlo samples.
We summarise the estimates as $\hat{S}^\diamond$ with $\diamond \in \lbrace i, Ti, ij \rbrace$ for the first, total, or second-order index estimates, respectively.
Remember that we now evaluate the Monte--Carlo samples with the metamodel instead of the full model.
We can, therefore, afford considerably larger numbers of Monte--Carlo samples.
Since we sample realisations of the GP metamodel $\GPsurrogate$, the resulting estimates $\hat{S}^\diamond$ are again random variables.
These include two sources of uncertainty: one related to the metamodel approximation and one related to the Monte--Carlo integration.
To estimate those uncertainties, and additionally the total uncertainty, we employ the algorithm suggested by~\cite{LeGratiet2014}.
The steps described in the following can equally be applied to all indices of different order.

We visually summarise the approach in Algorithm~\ref{algorithm1};
\begin{myalgorithm}
    \caption{\color{Gray} \sffamily \textbf{Calculation of {\normalfont $(\NGP \times B)$} estimates $\hat{S}_{k,b}^\diamond$ of the Sobol index}. We sample $\NGP$ realisations of the GP metamodel and subsequently resample each realisation $B$ times using the bootstrap technique as suggested by Le Gratiet \textit{et al.}~\cite{LeGratiet2014}.}
    \label{algorithm1}
    \includegraphics[width=0.98\linewidth]{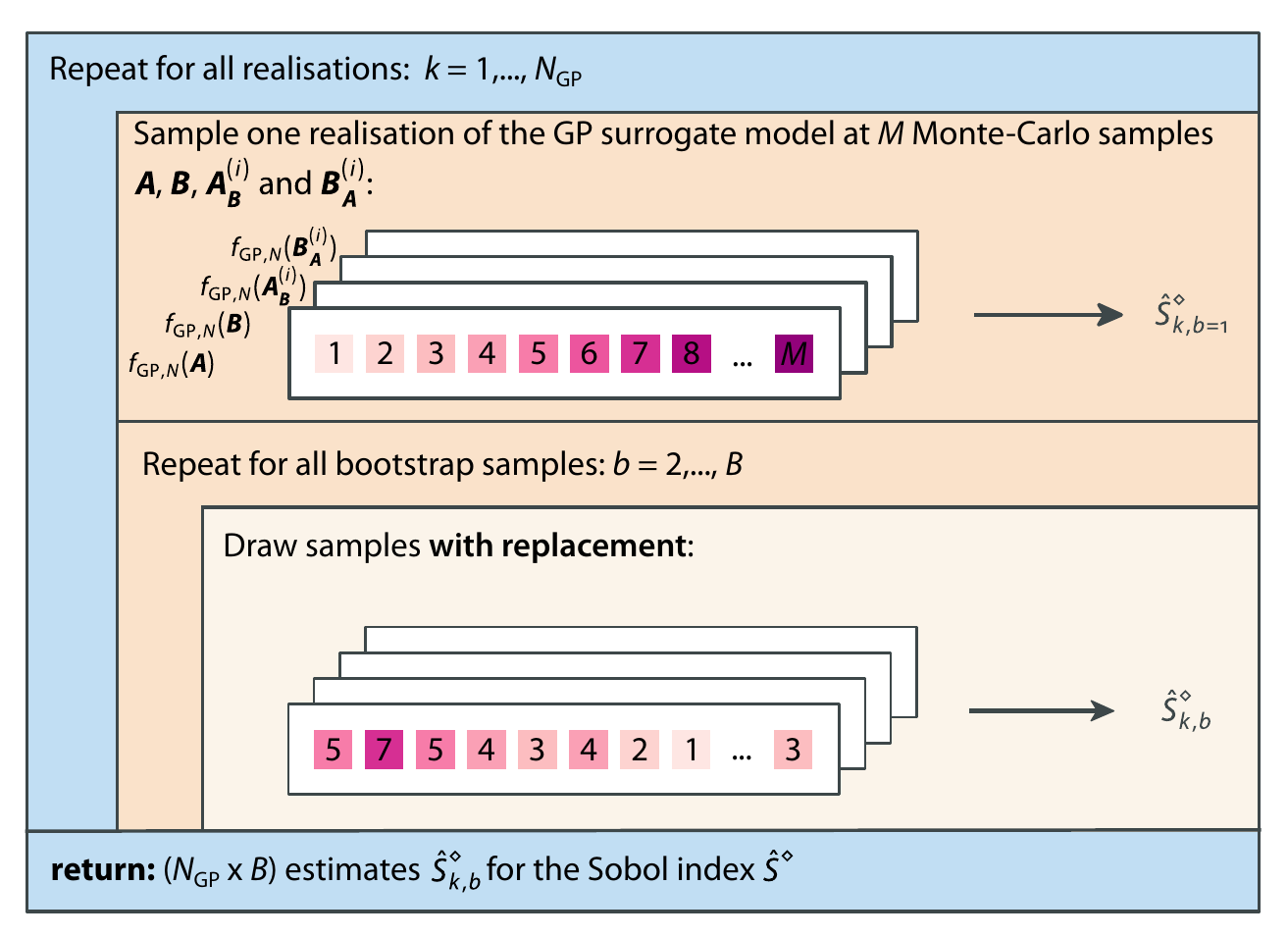}
\end{myalgorithm}a more detailed version is included in the Supplement A.4.
The core idea is to sample $\NGP$ realisations of the GP metamodel and, subsequently, resample each realisation $B$ times using the bootstrap technique \cite{Efron1979}.
This results in $\NGP \times B$ estimates $\hat{S}_{k,b}^\diamond$ of the respective Sobol index.
We then calculate the mean as
\begin{equation}\label{Eq:SobolEstimatorMean}
    \bar{S}^\diamond = \frac{1}{\NGP B}
    \sum\limits_{k = 1}^{\NGP}
    \sum\limits_{b = 1}^{B}
    \hat{S}_{k,b}^\diamond
\end{equation}
and the total variance as
\begin{equation}
    \hat{\sigma}^2 (S^\diamond)
    =
    \frac{1}{\NGP B - 1}
    \sum\limits_{k = 1}^{\NGP}
    \sum\limits_{b = 1}^{B}
    \left( \hat{S}_{k,b}^\diamond - \bar{S}^\diamond \right)^2.
\end{equation}
Since this estimator includes two sources of uncertainty (one related to the metamodel approximation and one related to the Monte--Carlo integration), we decompose the variance of $S^\diamond$ as
\begin{equation}
    \hat{\sigma}^2 (S^\diamond)
    =
    \underbrace{\qquad \hat{\sigma}_{\text{GP}}^2 (S^\diamond) \qquad}_{\text{metamodel}}
    \quad + \quad
    \underbrace{\qquad \hat{\sigma}_{\text{MC}}^2 (S^\diamond) \qquad}_{\text{Monte--Carlo}}.
\end{equation}

Sampling realisations of the metamodel $\GPsurrogate(\vct{X})$ as opposed to using only the predictive mean $\GPmean(\vct{X})$ allows us to take into account the covariance structure of the metamodel.
The part of the variance related to the metamodel approximation can be estimated as
\begin{equation}
    \label{Eq:VarianceMetamodel}
    \hat{\sigma}_{\text{GP}}^2 (S^\diamond)
    =
    \frac{1}{B}
    \sum\limits_{b=1}^{B}
    \frac{1}{\NGP - 1}
    \sum\limits_{k=1}^{\NGP}
    \left(
    \hat{S}_{k,b}^\diamond - \bar{\hat{S}}_{b}^\diamond
    \right)^2
    \hspace*{5pt}\text{with}\hspace*{5pt}
    \bar{\hat{S}}_{b}^\diamond
    =
    \frac{1}{\NGP}
    \sum_{k=1}^{\NGP} \hat{S}_{k,b}^\diamond.
\end{equation}
Alternatively, \cite{Janon2014} present an approach
to estimate an upper bound for the metamodel error based directly on the covariance function of the GP, but their approach only provides a rough upper bound \cite{LeGratiet2014}.
In addition, \cite{Panin2021} present an approach to investigate the accuracy of Sobol indices based on a general relation between the accuracy of an arbitrary metamodel and the error of the estimated indices.

To estimate the uncertainty related to Monte--Carlo integration, we use the bootstrap technique \cite{Efron1979}.
The Monte--Carlo samples $\Asaltellimatrix$, $\Bsaltellimatrix$, $\ABisaltellimatrix$ and $\BAisaltellimatrix$ are resampled (i.e. sampled with replacement) $B$ times as depicted in Algorithm~\ref{algorithm1}.
We then calculate $\hat{S}^\diamond$ according to \cref{Eq:SobolEstimatorFirstGP,,Eq:SobolEstimatorTotalGP,,Eq:SobolEstimatorSecondGP} for each bootstrap sample, resulting in $B$ estimates for the Sobol index for each realisation $k$ of the metamodel.
The part of the variance related to the Monte--Carlo integration is given by
\begin{equation}\label{Eq:VarianceMC}
    \hat{\sigma}_{\text{MC}}^2 (S^\diamond)
    =
    \frac{1}{\NGP}
    \sum\limits_{k=1}^{\NGP}
    \frac{1}{B - 1}
    \sum\limits_{b=1}^{B}
    \left(
    \hat{S}_{k,b}^\diamond - \bar{\hat{S}}_{k}^\diamond
    \right)^2
    \hspace*{5pt}\text{with}\hspace*{5pt}
    \bar{\hat{S}}_{k}^\diamond
    =
    \frac{1}{B}
    \sum_{b=1}^{B} \hat{S}_{k,b}^\diamond.
\end{equation}
The bootstrap technique is based on the fact that sampling with replacement from a set of independent, identically distributed data equals sampling from the empirical distribution function of the data \cite{Archer1997}.
It is important to note that bootstrapping does not require further model evaluations.
For a general introduction to the bootstrap technique, the reader is referred to \cite{Efron1993} or \cite[Sec. 5.2]{James2013}.

\section{Application to nanoparticle-mediated drug delivery in a multiphase tumour-growth model}

\subsection{Model definition}

An excellent example for the proposed overall approach is nanoparticle-mediated drug delivery in a multiphase tumour-growth model as all challenging motivating arguments for our approach are present in this problem class, like complex costly models and a large number of parameters.
The tumour-growth model in its original form is based on the works \cite{Sciume2013, Sciume2014,Sciume2014a}.
\cite{Kremheller2018, Kremheller2019} extended the model to a five-phase model including the vasculature.
Finally, \cite{Wirthl2020} included and studied nanoparticle delivery.
We use the tumour-growth model as a precursor to generate physically plausible results of the tumour and its microenvironment.
Those results then serve as initial condition for the sensitivity analysis of nanoparticle-mediated drug delivery.

\subsubsection{Underlying tumour-growth model}

\begin{figure}[ptb]

    \includegraphics[width=\textwidth]{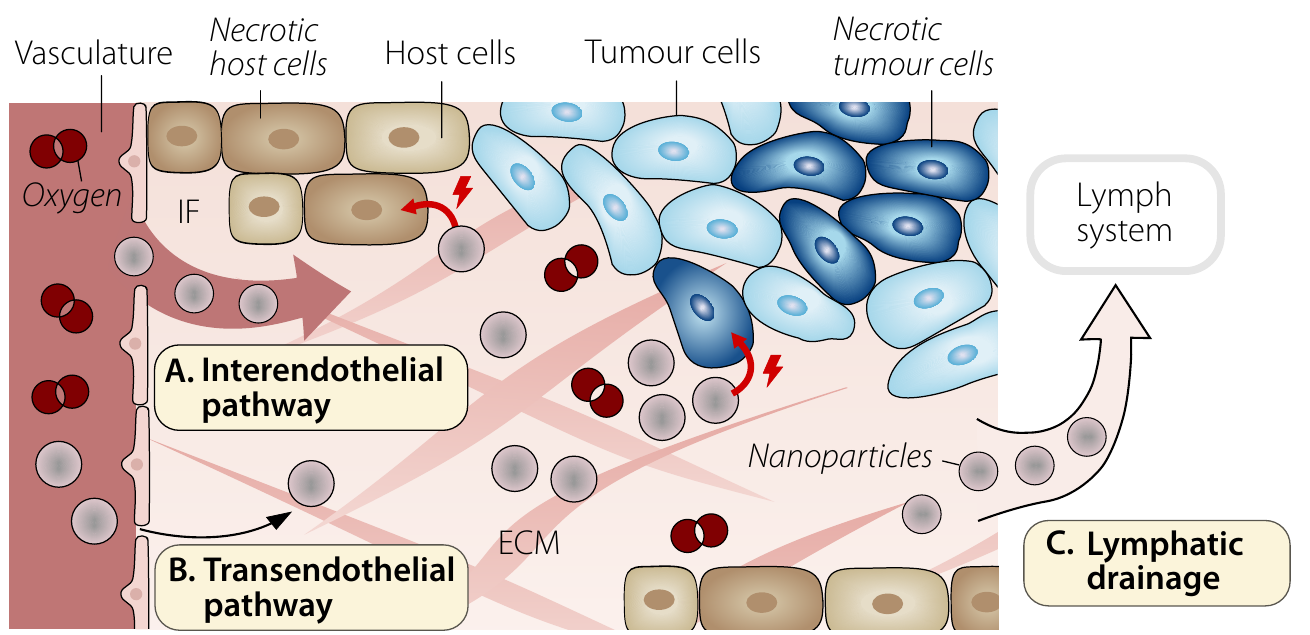}

    \caption{\color{Gray} \sffamily \textbf{Schematic summary of the components of the multiphase tumour-growth model}.
        The model comprises the ECM, as the solid phase, with three fluid phases filling its pore space: host cells, tumour cells and IF.
        The vasculature is included as an independent porous network.
        Species transported by the different phases are: oxygen, nanoparticles, necrotic host cells and necrotic tumour cells (marked in italics).
        We consider three mass transfer terms that transport nanoparticles into and out of the IF:
        A. the interendothelial pathway;
        B. the transendothelial pathway;
        C. lymphatic drainage
        (all transfer terms are marked in bold).
        Drugs mediated by the nanoparticles kill tumour and host cells (marked by red lightning).
        The lymph system is not explicitly included in the model.
    }
    \label{fig:TumourModelComponents}

\end{figure}

The model considers the tumour as a porous structure:
the extracellular matrix (ECM) is the solid phase (denoted by superscript $s$) with several fluid phases filling its pore space.
We include three fluid phases: tumour cells, host cells, and interstitial fluid (IF), denoted by superscripts $t$, $h$ and $l$, respectively.
In addition, the vasculature is modelled as an independent porous network and denoted by superscript $v$.
The governing equations of the model are formulated on the macroscale by employing the Thermodynamically Constrained Averaging Theory (TCAT) \cite{Gray2014,Miller2021}.
Each phase is modelled in an averaged sense based on volume fractions $\varepsilon^\alpha$ with $\alpha$ denoting an arbitrary phase.
The sum of all phases must satisfy the equation
\begin{equation*}
    \varepsilon^s + \varepsilon^t + \varepsilon^h + \varepsilon^l + \varepsilon^v = 1.
\end{equation*}
The different phases additionally transport species.
The vasculature and the IF transport oxygen with mass fractions denoted by $\oxygenv$ and $\oxygenIF$, respectively.
Similarly, $\NPv$ and $\NPl$ denote the mass fractions of nanoparticles in the vasculature and the IF, respectively.
Finally, tumour cells and host cells are divided up into living and necrotic cells.
The mass fraction of necrotic tumour cells is denoted by $\omega^{N\bar{t}}$ and the mass fraction of necrotic host cells by $\omega^{N\bar{h}}$.
\cref{fig:TumourModelComponents} schematically summarises all of the components of our multiphase tumour-growth model that are considered here.
Note that the lymph system is not explicitly modelled.

Many commonly used tumour-growth models are data driven and thus adhere to observed data.
Our model, by contrast, is based on physical laws.
To give just one example:
we use Fick's laws to describe the motion of oxygen in IF and Darcy's law to describe the flow of IF through the pores of the extracellular matrix.
Such a physics-based approach allows us to describe the system, even under unobserved circumstances.
Nevertheless, \cite[Sec. 1.1.4]{Saltelli2008} state that physics-based models are customarily over-parametrised:
they include more laws and parameters than available data would support.
This becomes particularly critical when model parameters are to be determined, e.g. by inverse analysis, and thus sensitivity analysis becomes a crucial part of model development \cite{Saltelli2020}.

\subsubsection{Nanoparticle-mediated drug delivery}
The transport of nanoparticles in our multiphase tumour-growth model is included as described in \cite{Wirthl2020}.
We only present a short summary in the following.
Further details can be found in the original publication \cite{Wirthl2020}.

Nanoparticles are intravenously injected and transported in the vasculature.
They then extravasate into the IF and are transported towards tumour cells and host cells.
The focus here lies on the extravasation into the IF and the transport therein.
Therefore, we do not explicitly include transport in the vasculature, but rather assume a constant mass fraction of nanoparticles in the vasculature $\NPv$.
The governing equation for the transport of nanoparticles in IF is the mass balance equation of nanoparticles with mass fraction $\NPl$
\begin{equation}
    \label{Eq:MassBalanceSpeciesTransport}
    \varepsilon^l \frac{\partial \NPl }{\partial t}
    -
    \underbrace{
        \frac{\mat{k}^l}{\mu^l} \vct{\nabla} p^l \cdot \vct{\nabla} \NPl
    }_
    {\substack{\textsf{convective} \\ \textsf{transport}}}
    -
    \underbrace{\phantom{\bigg(}
        \vct{\nabla} \cdot \left(\varepsilon^l \diffusivityNP \vct{\nabla} \omega^{\NP\bar{l}} \right)
    }_
    {\substack{\textsf{diffusive} \\ \textsf{transport}}}
    =
    \frac{1}{\rho^l}
    \bigg(
    \underbrace{
        \sum\limits_{\kappa} \overset{ \NP\kappa \rightarrow \NP l}{M}
    }_
    {\substack{\textsf{mass transfer} \\ \textsf{of NP to and from IF}}}
    - \hspace*{3pt}
    \NPl \sum\limits_{\kappa} \overset{\kappa \rightarrow l}{M}
    \bigg)
\end{equation}
where the effective diffusivity of nanoparticles in IF is given by $\diffusivityNP$.
The superscript $l$ denotes the IF as one of the fluid phases filling the pore space of the extracellular matrix.
The IF is characterised by its viscosity $\mu^l$, density $\rho^l$, permeability tensor $\mat{k}^l$, and finally the IF pressure $p^l$ resulting from the mass balance of the fluid equation as part of the tumour-growth model.
The last term results from employing the product rule, see \cite{Sciume2013,Kremheller2018}.
The mass transfer of nanoparticles to and from the IF includes three terms
\begin{equation}
    \sum\limits_{\kappa} \overset{\NP\kappa \rightarrow \NP l}{M}
    =
    \underbrace{\;\;
        \overset{\NP v \rightarrow \NP l}{M_{\textsf{inter}}}
        \;\;}_
    {\substack{\textsf{\vphantom{g}interendothelial}}}
    \;\; + \;\;
    \underbrace{\;\;
        \overset{\NP v \rightarrow \NP l}{M_{\textsf{trans}}}
        \;\;}_
    {\substack{\textsf{\vphantom{g}transendothelial}}}
    \;\; - \;\;
    \underbrace{\;\;
        \overset{\NP l \rightarrow \mathsf{NP}\hspace*{0.08em}\mathsf{ly}}{M_{\textsf{drain}}}
        \;\;}_
    {\substack{\textsf{drainage}}}
\end{equation}
where the physical interpretation of the different transport mechanisms is included in \cref{fig:TumourModelComponents}.
Nanoparticles extravasate from the vasculature into the IF through two different pathways: the interendothelial and the transendothelial pathway \cite{Wilhelm2016,Moghimi2018}.
Those pathways are described by
\begin{equation}
    \label{Eq:KedemKatchalsky}
    \begin{split}
        \overset{\mathsf{NP}v \rightarrow \mathsf{NP}l}{M}
        &=
        \overset{\mathsf{NP}v \rightarrow \mathsf{NP}l}{M_{\textsf{inter}}}
        \quad + \quad
        \overset{\mathsf{NP}v \rightarrow \mathsf{NP}l}{M_{\textsf{trans}}} \\
        &=
        \underbrace{
            \rho^v \varepsilon^v L_p^v  \frac{S}{V} \left[p^v - p^l - \sigma \left( \pi^v - \pi^l \right) \right] \frac{\NPv + \NPl}{2}
        }_
        {\substack{\textsf{interendothelial} \\ \textsf{pathway}}}
        \; + \;
        \underbrace{
            \rho^v \varepsilon^v P^{v} \frac{S}{V} \left\langle \NPv - \NPl \right\rangle_+
        }_
        {\substack{\textsf{transendothelial} \\ \textsf{pathway}}}
    \end{split}
\end{equation}
with the oncotic pressure difference between blood vessels and IF $\sigma \left( \pi^v - \pi^l \right)$, the surface-to-volume ration $S/V$, and Macaulay brackets $\langle \cdot \rangle_+$.
This equation is based on the Staverman--Kedem--Katchalsky equation similar to \cite{Jain1987}.

The first term describes the interendothelial pathway, which is a convective process:
nanoparticles are dragged by the transvascular fluid flow through gaps in the blood-vessel wall \cite{Jain1987}.
This process is governed by the hydraulic conductivity $L^v_p$ of the blood-vessel wall which is defined as
\begin{equation}
    \label{Eq:HydraulicConductivity}
    L^v_p = \frac{\gamma_p \, r_0^2}{8 \mu^v t}
\end{equation}
with the pore radius $r_0$, the vessel-wall thickness $t$, and the fraction of pores $\gamma_p$ \cite{Stylianopoulos2013}.

The second term in \cref{Eq:KedemKatchalsky} describes the transendothelial pathway, which is a diffusive process:
nanoparticles diffuse through the vessel-wall.
This diffusive flux is governed by the vascular permeability $P^v$.

Finally, the lymph system absorbs nanoparticles from the IF
\vspace*{-2mm}
\begin{equation}
    \overset{\mathsf{NP}l \rightarrow \mathsf{NP}\hspace*{0.08em}\mathsf{ly}}{M_{\textsf{drain}}}
    =
    \rho^l \NPl \lymphaticfiltration \left\langle p^l \right\rangle_+ \left\langle 1 - \frac{p^t}{p^{\mathsf{ly}}_{\mathsf{coll}}} \right\rangle _+
\end{equation}governed by the lymphatic filtration coefficient $\lymphaticfiltration$.
Lymphatic drainage is impaired above the collapsing pressure $p^{\mathsf{ly}}_{\mathsf{coll}}$, and thus no particles are removed.

We further assume that the nanoparticles transport and release anti-cancer drugs.
Those kill tumour cells and thus increase the mass fraction of necrotic tumour cells $\NTC$.
At the same time, those drugs have adverse side effects and kill host cells.
This additionally increases the mass fraction of necrotic host cells $\NHC$.
For the sake of simplicity, we assume that the mass fraction of killed cells is directly proportional to the mass fraction of nanoparticles present in the IF at a certain position.
We introduce intraphase reaction terms that increase the mass fraction of necrotic tumour cells and host cells according to
\begin{align}
    r_{\textsf{kill}}^{Nt} & = \gammaNTdrug \NPl (1 - \NTC) \label{Eq:KillingTumourCells} \\
    r_{\textsf{kill}}^{Nh} & = \gammaNHdrug \NPl (1 - \NHC) \label{Eq:KillingHostCells}
\end{align}
where $\gammaNTdrug$ and $\gammaNHdrug$ characterise the strength of the drug.

As quantity of interest for the sensitivity analysis, we consider the mean of the necrotic fraction of tumour cells given by
\begin{equation}
    \bar{\omega}^{N\bar{t}} = \frac{\int \omega^{N\bar{t}} \dd \Omega}{A^t}
\end{equation}
where the tumour size is defined as $A^t = \int \mathcal{H} (S^t - 0.1) \dd \Omega$ with the Heaviside function $\mathcal{H}(\cdot)$, and $S^t$ denotes the saturation of tumour cells \cite{Kremheller2018}.
We define the tumour as the part of the domain where $S^t > 0.1$.
In the context of sensitivity analysis, it is important to choose the quantity of interest carefully and to bear in mind that a parameter that is non-influential under one particular investigated condition, e.g., one particular quantity of interest, might be highly influential under a new condition \cite{Razavi2021}.

\subsection{Set-up of the numerical example}

The set-up presented in the following is similar to the example in \cite{Wirthl2020}.
We therefore only present a summary, and again refer the interested reader to the original publication for further details.
The major addition to the original example is the nanoparticle-mediated killing of tumour and host cells as described by \cref{Eq:KillingTumourCells,,Eq:KillingHostCells}.

We investigate nanoparticle transport and subsequent killing of cells by nano\-particle-mediated drugs.
The transport of nanoparticles depends on the hydraulic conductivity of blood-vessel walls $L_p^v$ and the blood-vessel wall permeability $P^v$ (both influence the transport of nanoparticles into the IF), the diffusivity $\diffusivityNP$ of nanoparticles in the IF, and the lymphatic filtration coefficient $\lymphaticfiltration$ (influencing the transport of nanoparticles out of the IF).
Subsequently, drugs mediated by the nanoparticles kill tumour and host cells depending on the strength of the drug, characterised by $\gammaNTdrug$ and $\gammaNHdrug$.
Note that the amount of killed cells largely depends on the amount of nanoparticles reaching a particular region of the domain, and hence depends on the transport parameters.

\begin{figure}[btp]
    \includegraphics[width=\textwidth]{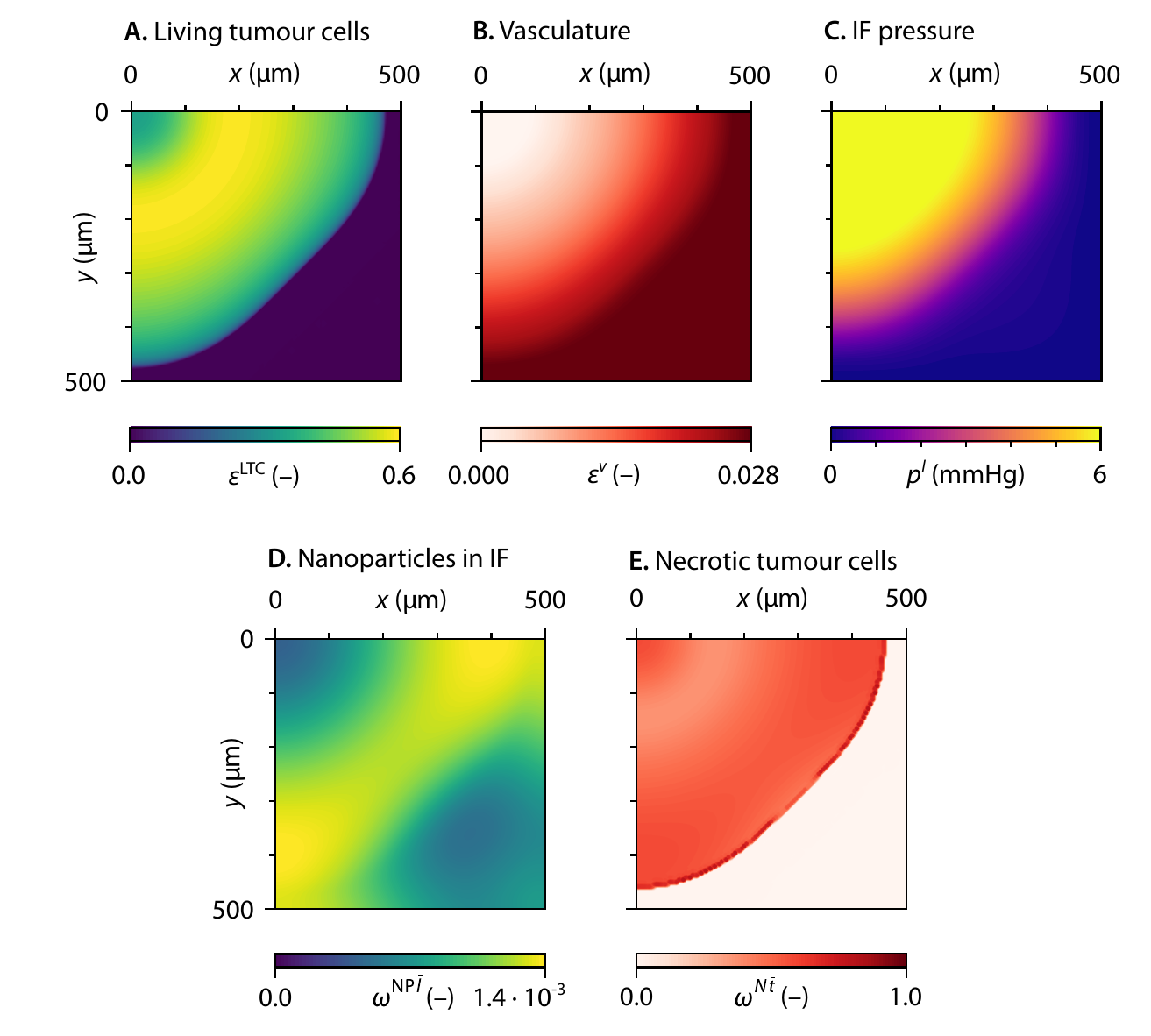}
    \caption{\color{Gray} \sffamily \textbf{Grown tumour in its microenvironment (A, B, C) with distribution of nanoparticles (D) and necrotic tumour cells (E).}
        A. Volume fraction of living tumour cells $\varepsilon^{\textsf{LTC}}$;
        B. Volume fraction of vasculature $\varepsilon^v$;
        C. Pressure $p^l$ in the IF;
        D. Mass fraction of nanoparticles in the IF $\NPl$;
        E. Mass fraction of necrotic tumour cells $\NTC$;
        Subfigures D and E as an example present the result for the mean values of the uncertain input parameters.}
    \label{fig:Tumour_microenv}
\end{figure}

The transport of nanoparticles, and thus the question of which regions nanoparticles reach and where drugs can kill cells, essentially depends on the microenvironment of the tumour.
Solid tumours exhibit typical features relevant in this context:
the majority of living tumour cells is located in the tumour periphery, whereas the tumour core mainly consists of necrotic cells (see \cref{fig:Tumour_microenv}A).
In addition, \cref{fig:Tumour_microenv}B shows that the volume fraction of the vasculature is considerably lower in the tumour area because the growing tumour collapses blood-vessels.
The inner core of the tumour even contains no vessels at all.
Finally, the interstitial pressure in the tumour is increased and can reach \SI{6}{\mmHg} (see \cref{fig:Tumour_microenv}C).
To sum up, the tumour has a necrotic core with collapsed blood vessels as well as an increased interstitial pressure, which is a structure typical for solid tumours and which has also been observed in experiments \cite{Cui2006,Macklin2009,Dewhirst2017,Stylianopoulos2012}.

We analyse a domain of $\SI{1}{\milli\meter} \times \SI{1}{\milli\meter}$, but due to the symmetry of the problem, we only simulate one quarter of the domain ($\SI{500}{\micro\meter} \times \SI{500}{\micro\meter}$).
The grown tumour has a radius of $\SI{440}{\micro\meter}$ as presented in \cref{fig:Tumour_microenv}.
We analyse a time interval of \SI{20}{\minute} of nanoparticle transport and killing of cells based on examples by \cite{Terentyuk2009, Nabil2015}.
Assuming the intravenous infusion of nanoparticles, we prescribe a constant value of $\omega^{\mathsf{NP}\bar{v}} = \num{2.0e-3}$ for the mass fraction of nanoparticles in the vasculature.

\cref{Tab:ParametersNanoparticles} summarises the six uncertain input parameters included in the sensitivity analysis.
We assume that all input parameters are distributed uniformly within the given ranges, which are based on experimental data (see references in \cref{Tab:ParametersNanoparticles}).
The uniform distribution is chosen because we lack more specific information about the input parameters:
given only the range of the input parameters (and no further information such as mean or variance), uniform distributions maximise the information entropy and hence minimise the introduced bias \cite{Shannon1948, Kullback1997}.
Note that the killing coefficient of host cells $\gammaNHdrug$ has no influence on our quantity of interest, the mean of the necrotic fraction of tumour cells---neither directly nor indirectly through coupling terms.
We nevertheless include the killing coefficient of host cells $\gammaNHdrug$ in the sensitivity analysis to investigate how reliably we can identify a non-influential input parameter as such.

    {
        \setlength{\aboverulesep}{0pt}
        \setlength{\belowrulesep}{0pt}
        \setlength{\extrarowheight}{.45ex}
        \begin{table}[tbp]
            \sffamily \small
            \caption{\color{Gray} \sffamily \textbf{Probability distributions of uncertain input parameters.} We assume that all parameters are distributed uniformly within the given range. }
            \label{Tab:ParametersNanoparticles}
            \begin{threeparttable}
                \begin{tabularx}{\linewidth}{l X l l l}
                    \midrule
                    \textbf{Symbol}              & \textbf{Parameter}                                         & \textbf{Range}                    & \textbf{Units}                                               & \textbf{Source}           \\
                    \rowcolor{naturelight}
                    $L_p^v$                      & Hydraulic conductivity of blood-vessel wall\tnote{$\star$} & $[7.8;\; 125] \cdot 10^{-8}$      & \si[per-mode=symbol]{\milli\meter\per\pascal\per \second}    & \cite{Stylianopoulos2013} \\
                    \rowcolor{white}
                    $P^v$                        & Blood-vessel wall permeability                             &
                    $[3.2;\; 128] \cdot 10^{-5}$ & \si[per-mode=symbol]{\milli\meter\per\second}              & \cite{Ho2017,Dreher2006,Chou2013}                                                                                            \\
                    \rowcolor{naturelight}
                    $\diffusivityNP$             & Diffusivity of nanoparticles                               & $[0.26;\; 30.83]$                 & \si[per-mode=symbol]{\micro\meter\cubed\per\second}          & \cite{Chou2013}           \\
                    \rowcolor{white}
                    $\lymphaticfiltration$       & Lymphatic filtration coefficient                           & $[0;\; 5.2] \cdot 10^{-4}$        & \si{\per\pascal\per\second}                                  & \cite{Wirthl2020}         \\
                    \rowcolor{naturelight}
                    $\gammaNTdrug$               & Killing coefficient of tumour cells                        & $[5;\; 10] \cdot 10^{-4}$         & \si[per-mode=symbol]{\gram\per\milli\meter\cubed\per\second} & --                        \\
                    $\gammaNHdrug$               & Killing coefficient of host cells                          & $[2;\; 7] \cdot 10^{-4}$          & \si[per-mode=symbol]{\gram\per\milli\meter\cubed\per\second} & --                        \\
                    \midrule
                \end{tabularx}
                \begin{tablenotes}
                    \footnotesize
                    \item[{($\star$)}] The given values for the hydraulic conductivity of the blood-vessel wall correspond to a pore radius of $r_0 = [50; \; 200]\;\si{\nano\meter}$ as used in \cite{Wirthl2020}.
                \end{tablenotes}
            \end{threeparttable}
        \end{table}
    }

\cref{fig:Tumour_microenv}D and E present a result for the distribution of nanoparticles in the IF and for the mass fraction of necrotic tumour cells:
for this example, we used the mean values of the six uncertain input parameters given in \cref{Tab:ParametersNanoparticles}.
Most nanoparticles accumulate at the edge of the tumour, while lymphatic drainage removes most particles outside the tumour area, and roughly 50\% of the tumour cells are necrotic.

The nanoparticle-mediated transport included in the multiphase tumour-growth model is implemented in our in-house research code BACI \cite{BACI2021}.
The sensitivity analysis methods, as presented above, are implemented in QUEENS \cite{Biehler2020}.
QUEENS is a general purpose framework for uncertainty quantification, physics-informed machine learning, Bayesian optimisation, inverse problems and simulation analytics on distributed computer systems.
We use GPy \cite{gpy2014} as a GP framework and PyTorch \cite{Paszke2019} to generate Sobol sequences.

\subsection{Predictive quality of the metamodel}

Since we use a GP metamodel to estimate the Sobol indices, we first assess its predictive quality.
To this end, we investigate the quality of the metamodel predictions for two different covariance functions $k(\GPx, \GPx')$ used for the GP:
we compare a tensorised, squared, exponential covariance function to a tensorised 5/2-Mat\'ern covariance function (with $\nu = 5/2$) \cite{Matern1986, Stein1999, Rasmussen2006}.
A tensorised covariance function has the form $k(\GPx, \GPx') = k_1(X_1, X_1') \cdot k_2(X_2, X_2') \cdot \ldots \cdot k_D(X_D, X_D')$ and as such includes a set of hyperparameters $\vct{\theta} = (\sigma_{fi}, \ell_i)_{i=1, \dots, D}$ for all input space dimensions, which we optimise by maximising the log marginal likelihood.

\begin{figure}[b!]
    \includegraphics[width=\textwidth]{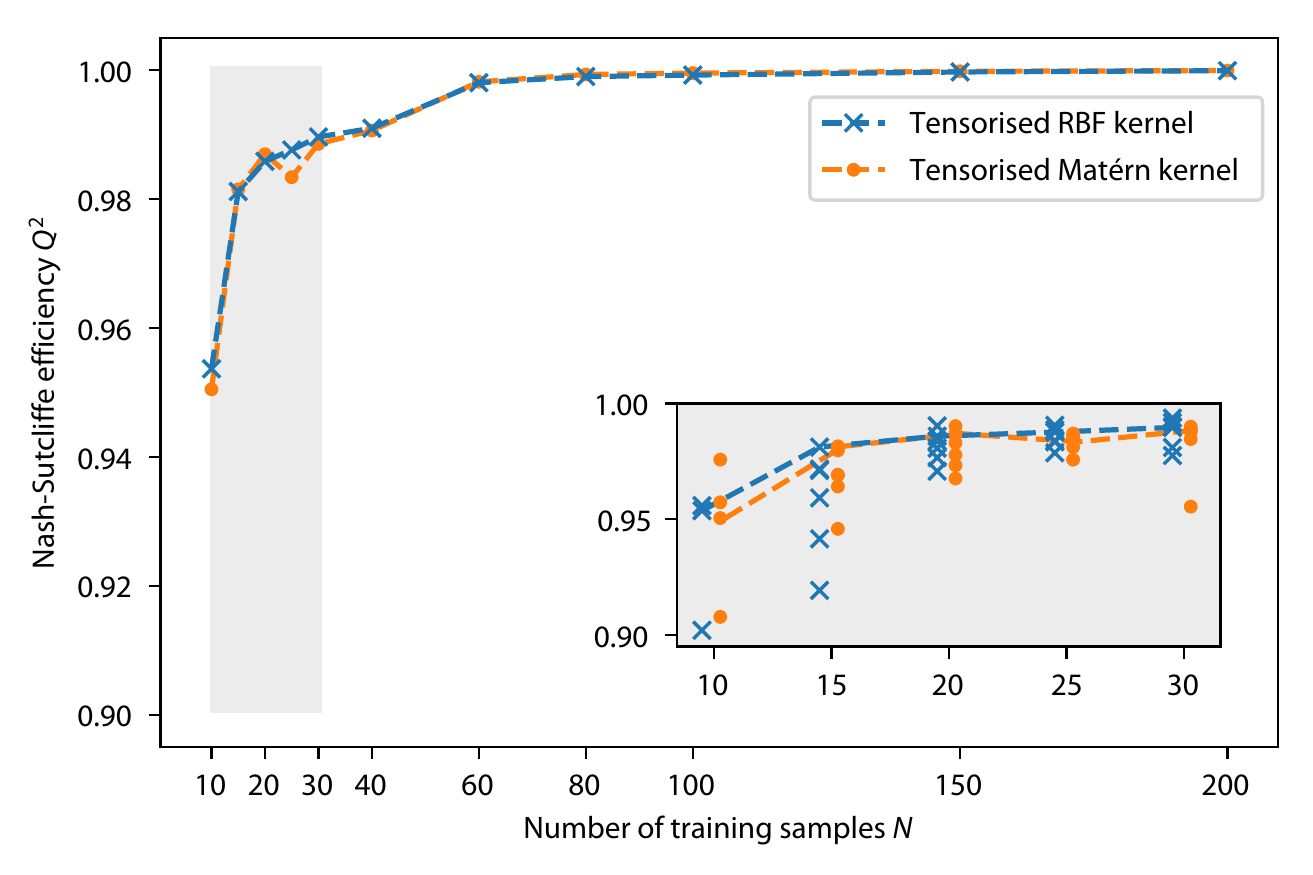}
    \caption{\color{Gray} \sffamily \textbf{Nash--Sutcliffe efficiency $Q^2$ for different training sample set sizes with a tensorised, squared, exponential covariance function and a tensorised Mat\'ern covariance function.}
    $N$ training samples were randomly generated for the main plot. For the detail plot, we repeated the process five times with different training sample sets of sizes $N = [10, 15, 20, 25, 30]$.
    For reference, the dashed lines in the detail plot are identical to the dashed lines in the main plot.}
    \label{fig:Predictive_quality}
\end{figure}

For this comparison, we consider different sizes of training sample sets
$N = [10, \allowbreak 15, \allowbreak 20, \allowbreak 25, \allowbreak 30, \allowbreak 40, \allowbreak 60, \allowbreak 80, \allowbreak 100, \allowbreak 150, \allowbreak 200]$,
which we generate based on Sobol sequences.
Additionally, we generate a set $\mathcal{T}$ of $N_T = 1000$ testing samples disjoint of the training samples.
Based on the training samples and the testing samples, we calculate the Nash--Sutcliffe efficiency $Q^2$ \cite{Nash1970} given by
\begin{equation}
    Q^2
    =
    1 -
    \frac{
        \sum_{\GPx \in \mathcal{T}} \left( \GPmean(\GPx) - f(\GPx) \right)^2
    }{
        \sum_{\GPx \in \mathcal{T}} \left( \GPmean(\GPx) - \bar{f}(\GPx) \right)^2
    }
    \quad \text{with} \quad
    \bar{f}(\GPx) =
    \frac{1}{T} \sum\limits_{\GPx \in \mathcal{T}} f(\GPx)
\end{equation}
similar to \cite{LeGratiet2014}.
This is based on the predictive posterior mean $\GPmean(\GPx)$ of the GP with optimised hyperparameters, and thus compares the mean prediction of the posterior GP to the actual output of the full model $f$.
A Nash--Sutcliffe efficiency close to $1$ indicates good agreement and, hence, reliable predictions.
\cref{fig:Predictive_quality} shows good convergence of the Nash--Sutcliffe efficiency for both covariance functions with values close to $1$, even for smaller training sample set sizes.
If the number of training samples is restricted due to the computational cost, \cite{VanSteenkiste2018} suggest an algorithm to improve the metamodelling accuracy and efficiency based on sequential sampling.

Note that we use a set of testing samples here that is disjoint of our set of training samples;
this means that we also evaluate our full model $N_T$ times, which might be infeasible if the model is computationally more expensive.
In those cases, one can use cross-validation methods, such as those explained in \cite[Sec. 5.3]{Rasmussen2006}, where the training set itself is split into two disjoint sets:
one is actually used for training and the other for validation.

Both covariance functions yield a very similar predictive quality.
In the following, we only use the tensorised, squared, exponential covariance function because it is the default choice in most applications of GPs \cite{Duvenaud2014} and, moreover, because it is a universal covariance function \cite{Micchelli2006}.

\begin{remark}[Randomness of metamodel training]
    In \cref{fig:Predictive_quality} we notice a small kink in the Nash--Sutcliffe efficiency for $N = 25$ in the case of the Mat\'ern covariance function.
    Therefore, we repeat the training of the GP with other randomly chosen training sample sets for $N = [10, 15, 20, 25, 30]$ to check whether the original training sets happen to perform exceptionally well.
    The grey detail plot in \cref{fig:Predictive_quality} presents the results:
    while the Nash--Sutcliffe efficiency is still above \num{0.90} in all cases, we notice that some training sample sets result in slightly worse efficiencies than others for those smaller sample sizes.
    This is precisely the case for the Mat\'ern covariance function with $N = 25$ in the main plot.
    Moreover, the two different covariance functions lead to slightly different efficiencies.

    One possible reason for such behaviour may be the convergence of the optimiser used to optimise the hyperparameters according to \cref{Eq:LogMarginalLikelihood}.
    As with all gradient-based optimisers, the optimisation may get stuck in a local minimum.
    To avoid ending the optimisation in a local minimum, one can repeat the optimisation multiple times from random initial points \cite{Tripathy2016}, as provided by the package GPy \cite{gpy2014}, for example, or use stochastic optimisation, such as Adam optimisation \cite{Kingma2017}.

\end{remark}

\begin{figure}[tbp]
    \includegraphics[width=\textwidth]{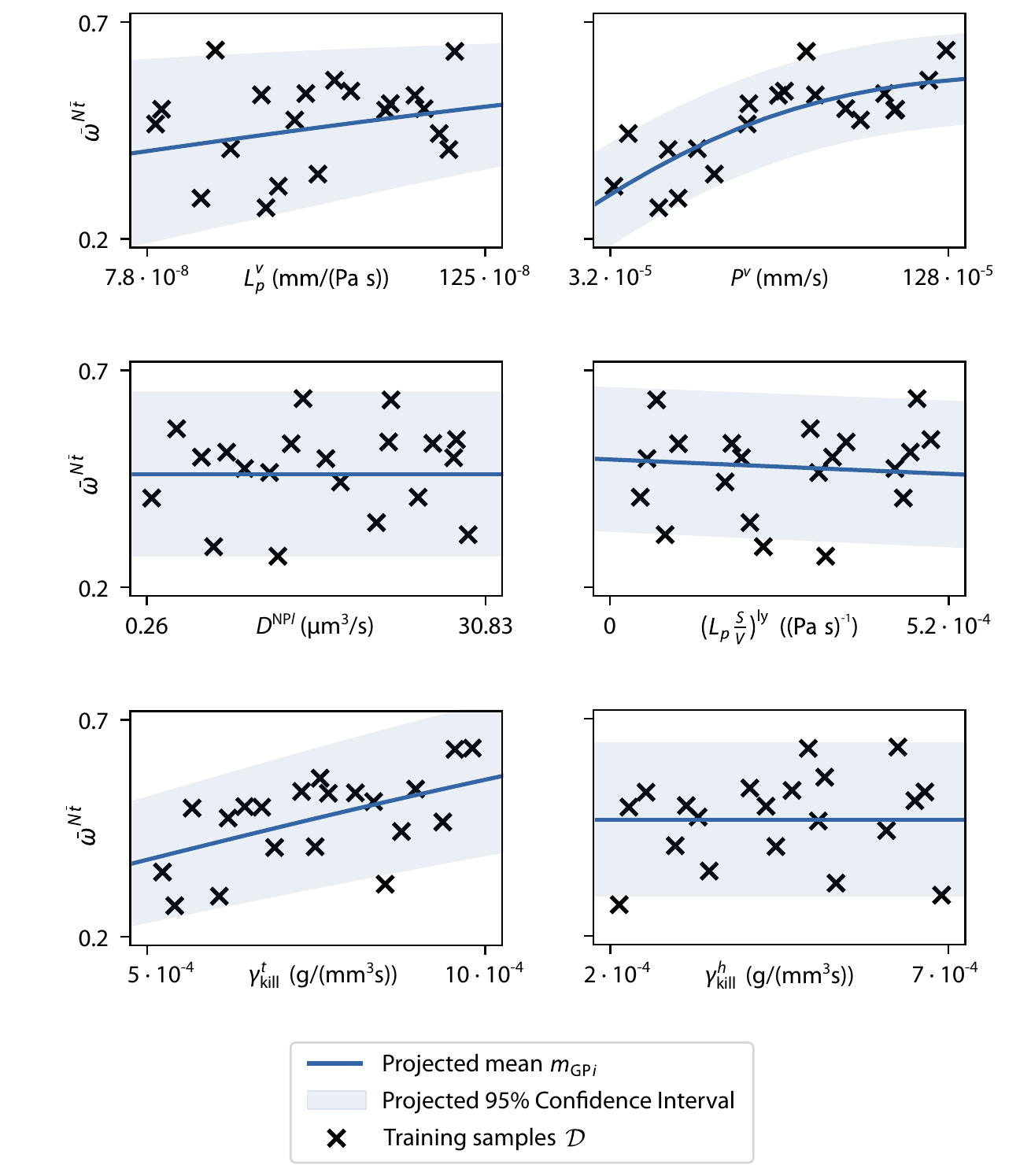}
    \caption{\color{Gray} \sffamily \textbf{Gaussian process for $N = 20$ with a tensorised, squared, exponential covariance function.} Projected mean $m_{\textsf{GP}i}(X_i)$, projected 95\% confidence interval (CI) and training samples $\datatrain$ for the mean of the necrotic fraction of tumour cells $\bar{\omega}^{N\bar{t}}$ (the y-axis labels apply to both figures).}
    \label{fig:GPs}
\end{figure}

Further, we take a closer look at the underlying GP, which is depicted in \cref{fig:GPs} for $20$ training samples.
The projection of the GP into the input-space dimensions reveals a linear relation in most dimensions.
We only see considerable nonlinearity for the blood-vessel wall permeability $P^v$.
Those characteristics make it much easier to train the GP based on a small number of training samples.

\begin{remark}[Projection of the $D$-dimensional Gaussian process]
    The projection $m_{\textsf{GP}i}(X_i)$ of the $D$-dimensional posterior mean $\GPmean(\GPx)$ into the input space dimension $X_i$ (as presented in \cref{fig:GPs}) is calculated as follows.
    First, we uniformly sample discrete values of the posterior mean $\GPmean(\GPx)$ in the $D$-dimensional input space.
    Second, we project those values over the input space dimension $X_i$.
    Third, the results are binned in the $X_i$-direction, and we calculate the mean and confidence interval for each bin.
    Note that we only project the posterior mean $\GPmean(\GPx)$ and neglect the covariance function $k_N$ here.

\end{remark}

Plotting the model output over a specific input in the form of scatterplots---as done with the training samples in \cref{fig:GPs}---helps us gain a general understanding of the magnitude of the underlying sensitivity \cite{Qian2020}.
\cite[Sec. 1.2.7]{Saltelli2008} offers a compelling interpretation of scatterplots in relation to the first-order Sobol index:
if the conditional expectation $\Ex{X_{\sim i}}{Y | X_i}$---here represented by the projection of the mean $m_{\textsf{GP}i}$ of the GP---has a large variation across $X_i$, the corresponding input parameter has a high first-order Sobol index.
\cref{fig:GPs} reveals that the projection of the mean $m_{\textsf{GP}i}$ is almost constant for the diffusivity of nanoparticles $\diffusivityNP$, the lymphatic filtration coefficient $\lymphaticfiltration$, and the killing coefficient of host cells $\gammaNHdrug$.
In contrast, the variation of the projection of the mean $m_{\textsf{GP}i}$ is larger for the blood-vessel wall $L_p^v$, the vascular permeability $P^v$, and the killing coefficient $\gammaNTdrug$, and we therefore expect the output to be highly sensitive to those parameters.

\subsection{First-order Sobol index estimates}

We now assess the convergence of the first-order Sobol index estimates for increasing numbers of training samples.
To calculate the mean $\bar{S}^i$ based on \cref{Eq:SobolEstimatorMean}, we use $\NGP = 500$ metamodel realisations, and the number of Monte--Carlo samples is set to $M = \num{10000}$.
We do not include bootstrapping here.
The uncertainties in the estimates will be studied in the next section.

\cref{fig:Estimator_comparison} presents the results for training sample set sizes $N = [10, \dots, 200]$.
The result confirms what we expected based on the scatterplots in the previous section:
three parameters---namely the vascular permeability $P^v$, the killing coefficient $\gammaNTdrug$, and the hydraulic conductivity of the blood-vessel wall $L_p^v$---have considerably higher first-order Sobol indices than the remaining three parameters.
\cref{fig:Estimator_comparison} also allows to assess the convergence of the Sobol indices for an increasing number of training samples:
even small sizes of training sample sets yield values close to the value based on $N = 200$.
This is due to the high values of the respective Nash--Sutcliffe efficiency, as discussed in the previous section.
Those results are promising, in particular for models that are computationally very expensive, and thus do not allow a large number of evaluations of the full model:
the computational cost of the demonstrated approach is considerably reduced compared to an analysis based directly on evaluations of the full model, as for example presented in \cite{Fritz2019,Phillips2020} for tumour-growth models.

We assessed convergence visually based on \cref{fig:Estimator_comparison}.
In addition, \cite{Sarrazin2016} present a thorough definition of convergence criteria for global sensitivity analysis results.
Nevertheless, the computationally limiting factor is usually the number of Monte--Carlo samples.
Since we evaluate the Monte--Carlo samples on the GP metamodel, this limitation is less critical in the presented workflow.
If sampling the realisations of the metamodel for very large numbers of Monte--Carlo samples becomes an issue, \cite{LeGratiet2014} include an efficient approach based on conditional GPs.

\subsection{Uncertainties of Sobol index estimation}

We now go on to not only estimate the mean $\bar{S}^i$ but also include the uncertainty related to the metamodel and to the Monte--Carlo integration given by \cref{Eq:VarianceMetamodel,,Eq:VarianceMC}.
In addition to the first-order index $S^i$, we also include the total-order Sobol index $S^{Ti}$.
Again, we use different training sample set sizes $N = [10, \dots, 200]$ and a tensorised, squared, exponential covariance function with hyperparameters optimised based on maximising the log marginal likelihood of the GP.
We draw $\NGP = 500$ realisations of the GP, $B = 300$ bootstrap samples, and $M = \num{10000}$ Monte--Carlo samples.
We calculate 95\% confidence intervals on the basis of the variance related to the metamodel $\hat{\sigma}^2_{\text{GP}}$ and the variance related to Monte--Carlo integration $\hat{\sigma}^2_{\text{MC}}$.

\cref{fig:Convergence_uncertainty}A presents the results for all six input parameters.
Note the different scaling on the  vertical axes.
We first take a look at the results for the indices themselves.
\cref{fig:Convergence_uncertainty}A again confirms that even for small numbers of training samples, the estimates for first and the total-order Sobol indices rapidly converge.
As mentioned above, we do not expect any influence of the parameter $\gammaNHdrug$ on the quantity of interest.
\cref{fig:Convergence_uncertainty}A shows that we can identify this non-influential parameter as such even for small numbers of training samples.
The parameter $\diffusivityNP$ also leads to Sobol indices close to zero for $N < 60$.
For larger training sample set sizes however we get a slightly higher total-order Sobol index, which is nevertheless small.
Hence, we can clearly separate the three most influential parameters from the three non-influential parameters.

Moreover, the total-order index is higher than the first-order index, in particular for the hydraulic conductivity of the blood-vessel wall $L_p^v$ and the vessel wall permeability $P^v$.
This leads to the conclusion that higher-order effects are indeed present.
We will therefore analyse the second-order Sobol indices in the next section.
In addition, \cref{Tab:FirstTotalOrderEffects} summarises the values for the first and the total-order indices for $200$ training samples:
the sum of all first-order Sobol indices is $0.967$.
Since this is close to one, we conclude that higher-order effects are present, but only play a minor role.
The largest part of the output variance is covered by the first-order indices.

    {
        \setlength{\aboverulesep}{0pt}
        \setlength{\belowrulesep}{0pt}
        \setlength{\extrarowheight}{.45ex}
        \begin{figure}[ptb]
            \includegraphics[width=\textwidth]{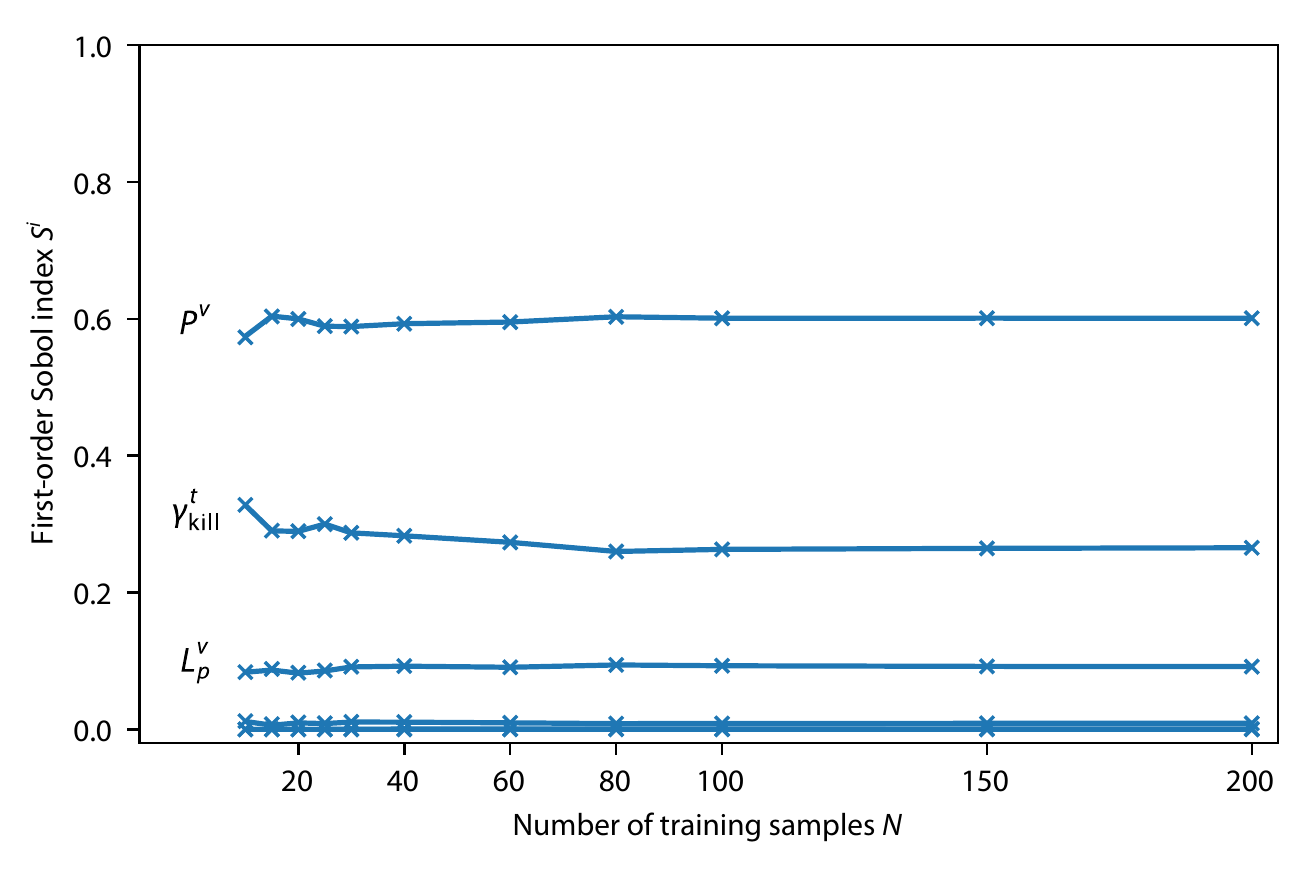}
            \caption{\color{Gray} \sffamily \textbf{Convergence of first-order Sobol index estimate for increasing training sample set size}.
                We use $M = \num{10000}$ Monte--Carlo samples and $\NGP = 500$ realisations of the Gaussian process.}
            \label{fig:Estimator_comparison}

            \vspace{3mm}
            \sffamily \small
            \centering
            \captionof{table}{\color{Gray} \sffamily \textbf{First and total-order Sobol indices based on 200 training samples.} We use $M = \num{10000}$ Monte--Carlo samples, $\NGP = 500$ realisations of the Gaussian process, and $B = 300$ bootstrap samples.}\label{Tab:FirstTotalOrderEffects}

            \begin{tabularx}{0.4\linewidth}{X l l}
                \midrule
                \textbf{Parameter}     & $S^i$                & $S^{Ti}$    \\
                \midrule
                \rowcolor{naturelight}
                $L_p^v$                & \num{0.092}          & \num{0.121} \\
                \rowcolor{white}
                $P^v$                  & \num{0.600}          & \num{0.632} \\
                \rowcolor{naturelight}
                $D^{\text{NP}l}$       & \num{0.001}          & \num{0.004} \\
                \rowcolor{white}
                $\lymphaticfiltration$ & \num{0.009}          & \num{0.011} \\
                \rowcolor{naturelight}
                $\gammaNTdrug$         & \num{0.265}          & \num{0.270} \\
                \rowcolor{white}
                $\gammaNHdrug$         & \num{0.000}          & \num{0.000} \\
                \rowcolor{naturedark}
                \textbf{sum}           & \textbf{\num{0.967}} &             \\
                \midrule
            \end{tabularx}
        \end{figure}
    }

We now focus on the uncertainties:
we assess the uncertainty related to the GP metamodel and the total uncertainty, where the latter includes both sources of uncertainty (related to Monte--Carlo integration and related to the metamodel).
For small training sample set sizes, we see considerable uncertainty related to the metamodel (depicted in light blue/orange in \cref{fig:Convergence_uncertainty}A).
\begin{figure}[ptb]
    \includegraphics[width=\textwidth]{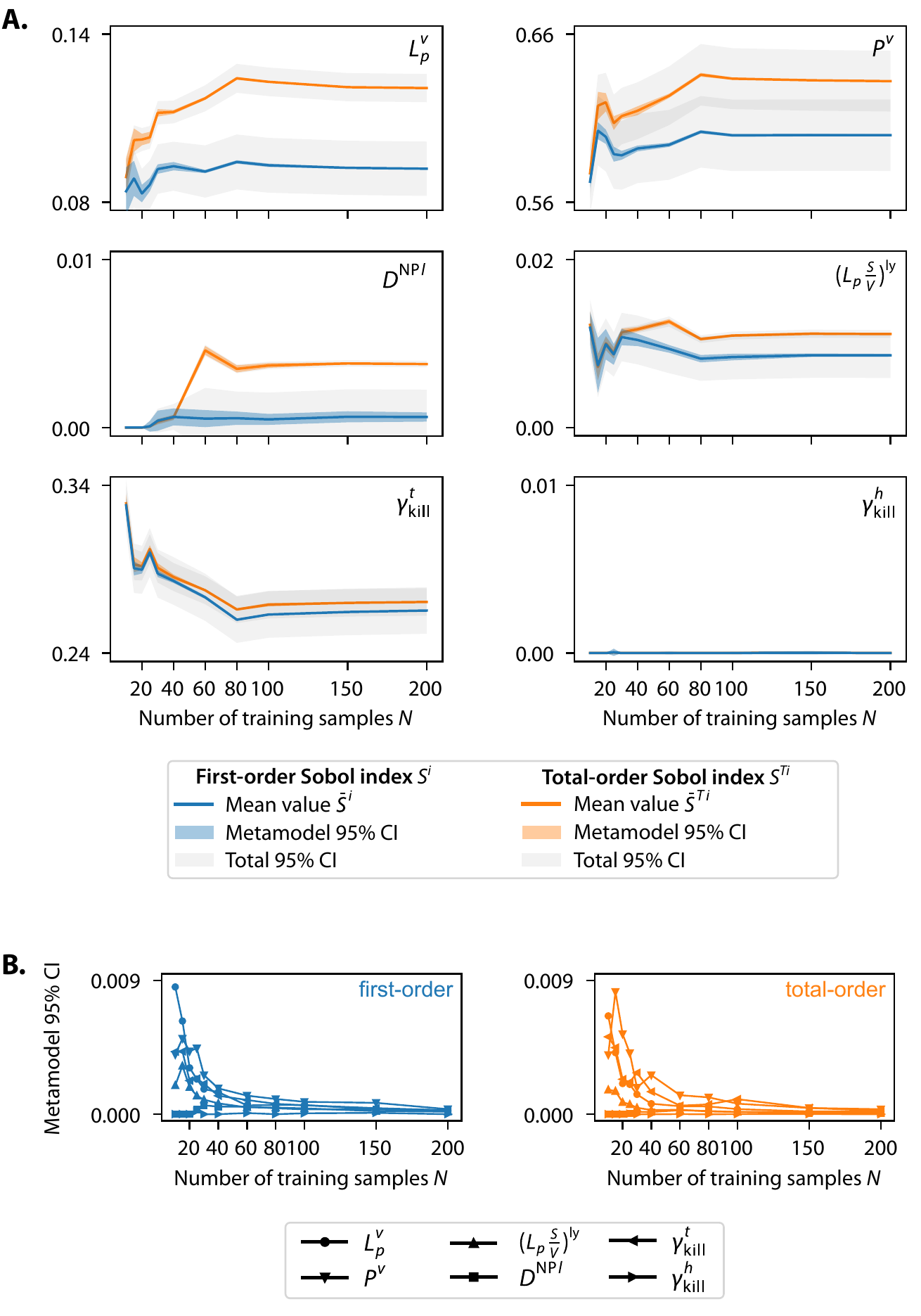}
    \caption{\color{Gray} \sffamily \textbf{First-order and total-order Sobol indices and 95\% confidence intervals (CI) for an increasing number of training samples}.
        We use $M = \num{10000}$ Monte--Carlo samples, $\NGP = 500$ realisations of the Gaussian process, and $B = 300$ bootstrap samples.
        Subfigure A shows Sobol indices with metamodel CI and the sum of metamodel and Monte--Carlo CI for the six input parameters separately.
        Monte--Carlo abbreviates as MC.
        Subfigure B details the metamodel CI.
    }
    \label{fig:Convergence_uncertainty}
\end{figure}
However, the uncertainty related to Monte--Carlo integration dominates for $N > 40$.
We therefore present the uncertainty related to the metamodel in detail in \cref{fig:Convergence_uncertainty}B:
the uncertainty rapidly decreases as the number of training samples increases for both the first-order and the total-order index and becomes one order of magnitude smaller than the uncertainty related to Monte--Carlo integration.
\cite{Hirvoas2021} also found the uncertainty related to the GP metamodel to be much smaller than the uncertainty related to Monte--Carlo integration in their example.
The total uncertainty (depicted in grey) could be reduced even further by increasing the number of Monte--Carlo samples.

Based on these results, we conclude that including the uncertainty related to the metamodel is not absolutely necessary in our example.
However, the example presented in the outlook and the example presented by \cite{LeGratiet2014} illustrate that this is not always the case:
only taking into account the Monte--Carlo uncertainty might then underestimate the confidence interval.
In such cases, it is essential to consider the uncertainty related to the metamodel.
Hence, this largely depends on the model, the input parameters, and the quantity of interest, and no \textit{one-size-fits-all} rule can be given.

Nevertheless, even taking into account the metamodel and the Monte--Carlo uncertainty may incorrectly estimate the confidence intervals:
poor optimisation of the hyperparameters may result in underestimated or overestimated confidence intervals.
In such cases, one could additionally consider the uncertainty related to the estimation of the hyperparameters of the GP covariance function by using a full-Bayesian approach with hyperpriors \cite{LeGratiet2014}.

To sum up, the demonstrated workflow not only identifies parameters with a high first-order Sobol index (necessary for factor prioritisation) but also parameters with a small total-order Sobol index (necessary for factor fixing).
In both cases, small numbers of training samples suffice in our example.

\subsection{Second-order Sobol index estimation}

Since we concluded from the results in the previous sections that higher-order effects are indeed present in our example, the goal now is to estimate the second-order Sobol indices, and thereby identify interaction effects between the input parameters.
The results in the previous section show that the uncertainty related to Monte--Carlo integration is dominant, and the uncertainty related to the GP metamodel is much smaller.
We therefore estimate the second-order indices based on the predictive mean $\GPmean(\Xvector)$ of the GP and do not take into account the uncertainty related to the metamodel.
Thus, we estimate the second-order Sobol indices as
\begin{equation}
    \hat{S}^{ij}
    =
    \frac{
        \frac{1}{M} \sum_{m=1}^{M}
        \GPmean(\BAisaltellimatrix)_m \; \GPmean(\ABjsaltellimatrix)_m
        -
        \GPmean(\Asaltellimatrix)_m \; \GPmean(\Bsaltellimatrix)_m
    }{
        \mathcal{V}[{\GPmean([\Asaltellimatrix\;  \Bsaltellimatrix])}]
    }
    -
    \hat{S}^i
    -
    \hat{S}^j
    \label{Eq:SobolEstimatorSecondGPmean}
\end{equation}
which includes estimating the first-order effects $\hat{S}^i$ and $\hat{S}^j$ also based on the mean of the GP only.

Using the same number of Monte--Carlo samples as before ($M = \num{10000}$), however, results in a 95\% confidence interval with the same order of magnitude as the indices themselves, and even leads to negative values for the Sobol indices (as given in \cref{Tab:SecondOrderEffects}).
Therefore, we increase the number of Monte--Carlo samples to $M = \num{1000000}$ to obtain reasonably small confidence intervals.

    {
        \setlength{\aboverulesep}{0pt}
        \setlength{\belowrulesep}{0pt}
        \setlength{\extrarowheight}{.45ex}

        \begin{figure}[ptb]
            \sffamily \small
            \centering
            \captionof{table}{\color{Gray} \sffamily \textbf{Second-order Sobol indices for $M = \num{1000000}$ Monte--Carlo samples and $B = 300$ bootstrap samples.} All other second-order indices are $S^{ij} < 0.001$.}\label{Tab:SecondOrderEffects}
            \begin{tabularx}{0.8\linewidth}{l l l l X l l}
                \midrule
                \textbf{Parameter} $i$          & \textbf{Parameter} $j$   & \multicolumn{2}{c}{$M = \num{10000}$} &              & \multicolumn{2}{c}{$M = \num{1000000}$}                               \\
                \cmidrule{3-4} \cmidrule{6-7}
                                                &                          & $S^{ij}$                              & 95\% CI      &                                         & $S^{ij}$     & 95\% CI      \\
                \rowcolor{naturelight}
                $L_p^v$                         & $P^v$                    & \phantom{$-$}\num{0.0276}             & \num{0.0198} &                                         & \num{0.0283} & \num{0.0021} \\
                \rowcolor{white}
                $L_p^v$                         & $D^{\text{NP}l}$         & \num{-0.0009}                         & \num{0.0136} &                                         & \num{0.0003} & \num{0.0016} \\
                \rowcolor{naturelight}
                $L_p^v$                         & $\gamma^t_{\text{kill}}$ & \num{-0.0004}                         & \num{0.0142} &                                         & \num{0.0007} & \num{0.0017} \\
                \rowcolor{white}
                $P^v$                           & $\gamma^t_{\text{kill}}$ & \phantom{$-$}\num{0.0036}             & \num{0.0312} &                                         & \num{0.0042} & \num{0.0030} \\
                \rowcolor{naturelight}
                $(L_p \frac{S}{V})^{\text{ly}}$ & $D^{\text{NP}l}$         & \phantom{$-$}\num{0.0021}             & \num{0.0037} &                                         & \num{0.0023} & \num{0.0004} \\
                \midrule
            \end{tabularx}

            \includegraphics[width=\textwidth]{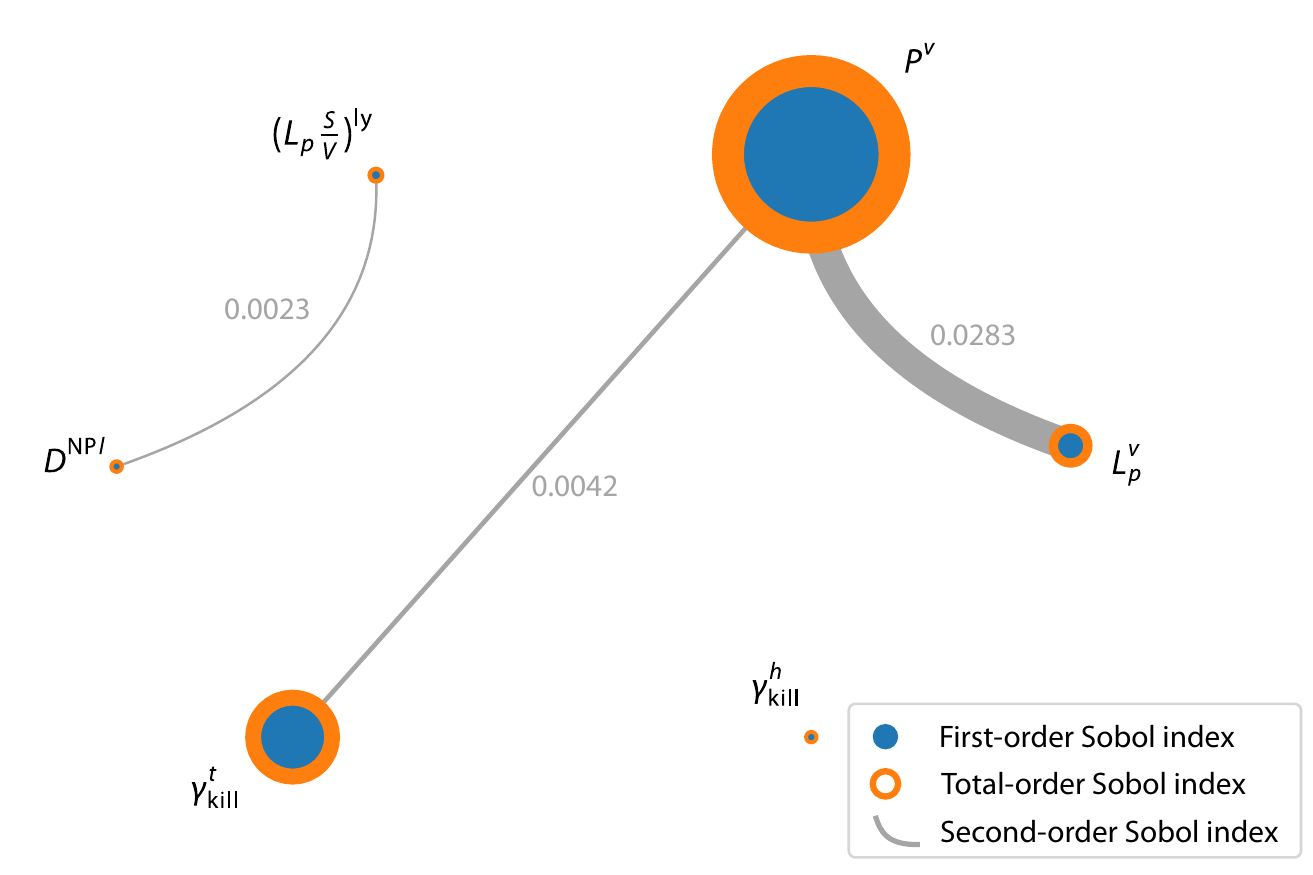}
            \captionof{figure}{\color{Gray} \sffamily \textbf{Second-order Sobol indices $\hat{S}^{ij}$ for $M = \num{1000000}$ Monte--Carlo samples and $B = 300$ bootstrap samples}. The blue circles represent the first-order Sobol indices. The orange surrounds represent the total-order Sobol indices. The grey areas connecting the nodes represent the second-order Sobol indices. All other second-order indices are $\hat{S}^{ij} < 0.001$.}\label{fig:Second_order}
        \end{figure}
    }

\cref{Tab:SecondOrderEffects,,fig:Second_order} summarise the results for the second-order Sobol indices:
the highest interaction is present between $P^v$ and $L_p^v$, as we already expected based on the results presented in \cref{fig:Convergence_uncertainty}.
Summing up all first and second-order Sobol indices results in $\num{0.999}$.
We thus (almost) completely apportioned the variance in the output to the input parameters, including interaction effects.

The large number of Monte--Carlo samples necessary to estimate the second-order Sobol indices highlights the relevance of metamodel-based estimation approaches.
Evaluating the full model $f$ several million times is computationally prohibitive for most models.
Without using a metamodel, estimating higher-order Sobol indices is thus impossible in most cases.

\section{Application to a model of arterial growth and remodelling}

In the previous sections, we assessed the performance of the current workflow as applied to a model of nanoparticle-mediated drug delivery in more detail:
even small numbers of training samples result in reliable estimates of the Sobol indices and a small uncertainty related to the GP metamodel.
To give an outlook, we now apply the workflow to another complex biomechanical example, namely a homogenised, constrained mixture model of arterial growth and remodelling \cite{Cyron2016,Cyron2017,Braeu2019}.

\cite{Brandstaeter2021} performed an exhaustive global sensitivity analysis, where they estimated the first and total-order Sobol indices by evaluating the full model for all Monte--Carlo samples.
This however entails a large computational burden (> \num{70000} model evaluations).
Therefore, the question arises as to whether we can reduce this computational cost by using a GP metamodel and still get reliable Sobol index estimates, including reliable uncertainty estimates.

For this comparison, we investigate \textit{Case 2} of the original publication \cite{Brandstaeter2021}, where the maximum diameter of an idealised cylindrical abdominal aorta was studied 15 years after spontaneous damage to elastin.
In this case, the majority of samples lead to minor dilatation of the vessel $\dmax < \SI{3}{\centi\meter}$.
In contrast, a considerable number of samples do not stabilise and keep enlarging, leading to aneurysms with a much larger diameter $\dmax >> \SI{3}{\centi\meter}$ (see Fig~4b in \cite{Brandstaeter2021}).
We use the original results from \cite{Brandstaeter2021} as a reference and compare them to our results based on the metamodel approach.

\begin{figure}[ptb]
    \includegraphics[width=\textwidth]{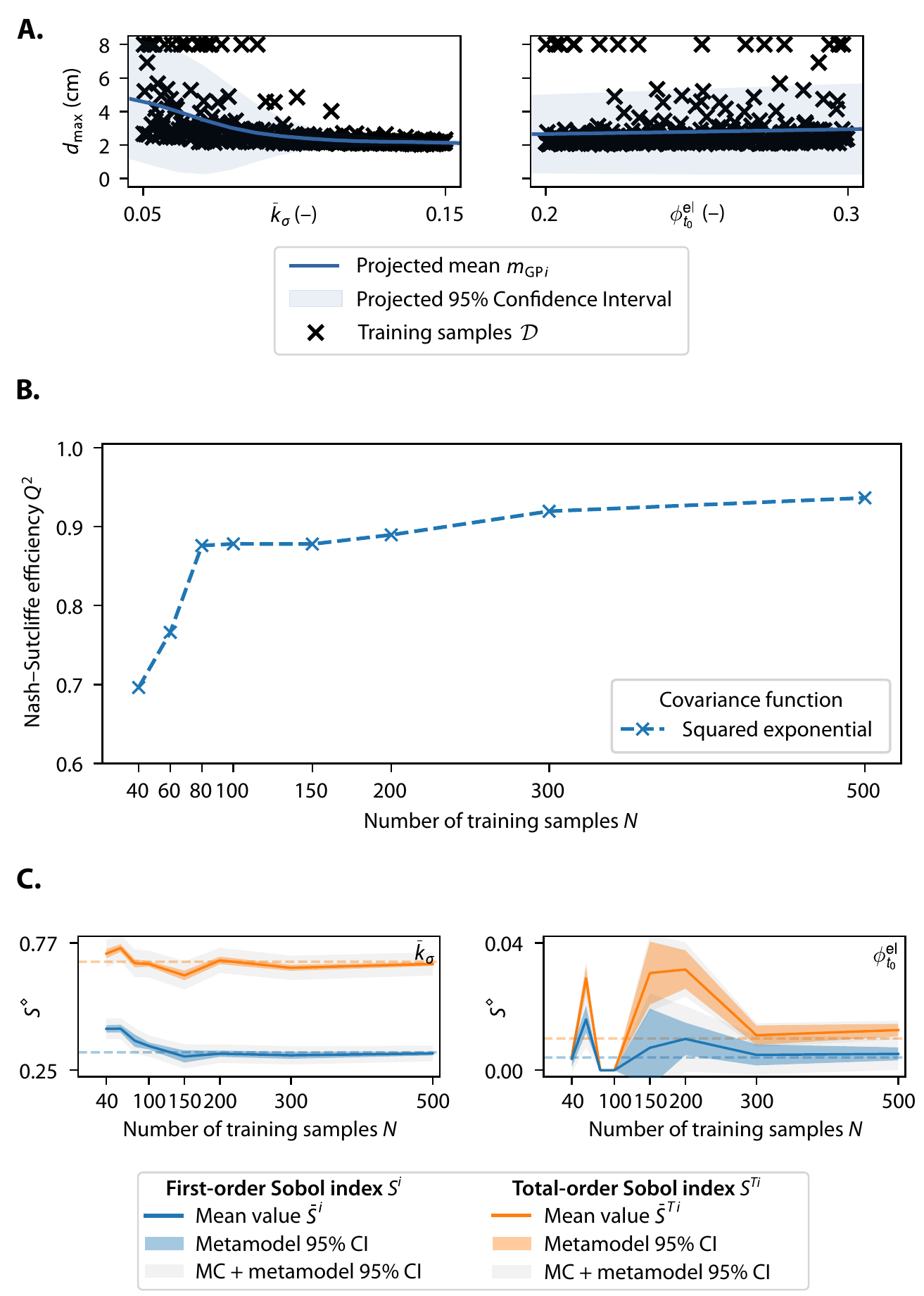}
    \caption{\color{Gray} \sffamily \textbf{Summary of results for arterial growth and remodelling}. A.  Projected mean $m_{\textsf{GP}i}(X_i)$, projected 95\% confidence interval (CI) and training samples $\datatrain$ for the diameter $\dmax$ (the y-axis labels apply to both figures).
        B. Nash--Sutcliffe efficiency $Q^2$ for different training sample set sizes.
        C. First-order and total-order Sobol indices and 95\% confidence intervals (CI) for an increasing number of training samples.
        Monte--Carlo abbreviates as MC.}
    \label{fig:AAA_Summary}
\end{figure}

First, we take a look at the predictive quality of the GP metamodel.
Once again, we use a tensorised, squared, exponential covariance function and compare the results for different numbers of training samples, $N = [40, 60, 80, 100, 150, 200, 300, 500]$ in this case.
As an example, \cref{fig:AAA_Summary}A presents the training samples for $N = 300$ for two parameters.
The results for the remaining parameters are included in the Supplement (see Fig~A.1 in Supplement).
We see that the majority of samples result in a small dilatation in contrast to the fewer aneurysmatic samples with a very large diameter of up to \SI{8}{\centi\meter}.
This bimodal structure of the data makes training the GP metamodel more difficult compared to our previous example.
Accordingly, the Nash--Sutcliffe presented in \cref{fig:AAA_Summary}B is lower, particularly for small numbers of training samples, i.e. $N < 80$.

Second, we calculate the first and total-order Sobol indices and respective uncertainties for different numbers of training samples.
To this end, we use $M = \num{10000}$ Monte--Carlo samples, $\NGP = 500$ realisations of the GP metamodel, and $B = 300$ bootstrap samples.
By way of example, we present the results for two parameters, the gain parameter $\bar{k}_\sigma$ and the initial volume fraction of elastin $\phi_{t_0}^{\textsf{el}}$, in \cref{fig:AAA_Summary}C.
Similar plots for the remaining eight parameters are included in the Supplement (see Fig.~A.2 in Supplement).
For the gain parameter $\bar{k}_\sigma$, the estimates based on the metamodel converge to the reference values from \cite{Brandstaeter2021} for both the first and the total-order Sobol index.
For the initial volume fraction of elastin $\phi_{t_0}^{\textsf{el}}$, the reference values for the Sobol indices are very small (0.01 or smaller).
In this case, exact estimates based on the metamodel approach are much harder to achieve:
for $N \leq 300$ training samples the estimates only stabilise.
Nevertheless, we can still reliably separate the three most influential parameters from the non-influential parameters, even for small numbers of training samples (see Fig.~A.2 in Supplement).
One further detail should be mentioned as an example:
the plot for $\phi_{t_0}^{\textsf{el}}$ reveals problems in estimating the first and total-order Sobol indices for $N = 80$ or $100$.
The Sobol indices and the uncertainties are all close to zero.
Similar behaviour can be observed for other parameters (see Fig.~A.2 in Supplement).
For $N < 300$, the GP has not yet converged, and hence does not capture all features of the quantity of interest.
Furthermore, we note that the uncertainty related to the metamodel is much higher and in some cases even dominates the total uncertainty in this example, while the uncertainty related to Monte--Carlo integration dominated the previous example.
It is therefore important to include the uncertainty related to the metamodel because considering only the uncertainty related to Monte--Carlo integration would underestimate the total uncertainty in the Sobol index estimate.

Finally, the computation of higher-order indices was not feasible with the approach chosen in the original contribution \cite{Brandstaeter2021}.
In contrast, the following will show that the metamodel-based approach enables their computation.
As an example, we again consider the gain parameter $\bar{k}_\sigma$:
a closer look at \cref{fig:AAA_Summary}C reveals that the total-index is considerably higher than the first-order index:
$S^{T\bar{k}_\sigma} - S^{\bar{k}_\sigma} = 0.34$.
This delta indicates interactions with other parameters, and thus estimating the second-order indices is of particular interest for this example.
Since we see in \cref{fig:AAA_Summary}C that the metamodel contributes significantly to the total uncertainty, we include uncertainty estimates for the metamodel (as opposed to relying solely on the predictive mean, as in the previous second-order estimates).
The estimates indeed show interaction with two parameters:
the turnover time $\tau$\footnotemark and the constitutive parameter $k_2$
($S^{\bar{k}_\sigma \tau} = 0.19$ and $S^{\bar{k}_\sigma k_2} = 0.06$).
However, the sum of all second-order indices ($S^{\bar{k}_\sigma j} = 0.25$) still does not cover the delta between the first and total-order index.
Hence, we specifically estimate the third-order Sobol index for the three most influential parameters, $\bar{k}_\sigma$, $\tau$, and $k_2$, resulting in considerable third-order interaction:
$S^{\bar{k}_\sigma \tau k_2} = 0.06$.
Detailed results for the second and third-order indices, including confidence intervals, are included in the Supplement (see Fig.~A.3 in Supplement).

\footnotetext{The turnover time was denoted by $T$ in the original publication \cite{Brandstaeter2021}.}

Thus, we are able to identify influential parameters for factor prioritisation based on the metamodel approach with small numbers of training samples, and we can also separate the influential parameters from the non-influential ones.
Hence, the metamodel-based approach provides the same results as the approach based directly on the full model in the original publication \cite{Brandstaeter2021}.
The computational cost, i.e., the number of evaluations of the full model, however, is much lower when using a metamodel.
Additionally, we can quantify higher-order indices which is infeasible based on evaluations of the full model.

\section{Conclusion}

Since a global sensitivity analysis is computationally expensive, modellers often rely on local methods alone, which may be inadequate \cite{Saltelli2019}.
The use of a metamodel-based approach, however, allows a global variance-based sensitivity analysis to be performed, even for computationally expensive biomechanical models with a moderate number of input space dimensions at a manageable computational cost.
The number of training samples required to obtain reliable estimates for the Sobol indices depends largely on the problem set-up itself:
our results demonstrate that we can identify the most influential input parameters and separate them from non-influential parameters with small numbers of training samples.
However, quantifying the exact value of the Sobol indices requires more training samples.
Moreover, the approach is able to quantify the uncertainty related to the metamodel:
including this uncertainty is important, because considering only the uncertainty related to Monte--Carlo integration could underestimate the total uncertainty in the Sobol index estimates.
The metamodel-based approach also allows an estimation of higher-order Sobol indices, and thus a quantification of interaction effects, which is not feasible without a metamodel due to the computational costs involved.
While there is no \textit{one-size-fits-all} rule, the approach is general and efficient enough to allow a study of different aspects of sensitivity analysis, including a transparent declaration of the uncertainties involved in the estimation process.

We demonstrated how a rigorous global sensitivity analysis can be applied to complex, computationally expensive problems.
A carefully performed sensitivity analysis is generally an integral part to ensure the high quality of any model development \cite{Saltelli2019}.
By demonstrating the workflow and its application for biomechanical problems, we contribute to closing the gap between proposals of new sensitivity analysis methods and application papers \cite{Razavi2021}.
We hereby encourage sensitivity analysis in general and the metamodel-based approach in particular in the biomechanics community.
In the big picture of model development, the presented workflow can be a building block towards inverse analysis, or it can be a valuable tool to better understand the model itself.

\section*{Acknowledgements}
SB wishes to acknowledge funding of the Deutsche Forschungsgemeinschaft (DFG, German Research Foundation) via project 257981274 and 386349077, JN and WAW wish to acknowledge funding of the Deutsche Forschungsgemeinschaft (DFG, German Research Foundation) via project WA 1521/23, and BAS gratefully acknowledges the support of the Institute for Advanced Study -- Technical University of Munich.
The original version of the software QUEENS was provided by the courtesy of AdCo Engineering\textsuperscript{GW} GmbH, which is gratefully acknowledged.

\printbibliography



\section*{Supplementary material}

# SUPPORTING INFORMATION

# Global sensitivity analysis based on Gaussian-process metamodelling for complex biomechanical problems

Barbara Wirthl, Sebastian Brandstaeter, Jonas Nitzler, Bernhard A. Schrefler, Wolfgang A. Wall

## A.1 Expected value and variance

We have a deterministic (forward) model $f$ so that

$$y = f(\boldsymbol{x}). \tag{A.1}$$

For every set of input parameters $\boldsymbol{x}$, the model returns the corresponding output $y$. We assume the model to be deterministic, i.e., the model itself does not introduce any uncertainty. In contrast, the inputs are uncertain, and we represent them as random variables $X_i$ summarised in a random vector $\boldsymbol{X}$ described by a probability density function $p(\boldsymbol{x})$. Then, the response random variable $Y$ is defined by

$$Y = f(\boldsymbol{X}) \tag{A.2}$$

with the expected value (first moment) given by

$$\mathbb{E}\left[Y\right] = \int Y \, p(\boldsymbol{x}) \, \mathrm{d}\boldsymbol{x} \tag{A.3}$$

and the variance (second moment) given by

$$\sigma_Y^2 = \mathbb{V}\left[Y\right] = \int \left(Y - \mathbb{E}\left[Y\right]\right)^2 p(\boldsymbol{x}) \, \mathrm{d}\boldsymbol{x} \tag{A.4}$$

where $\sigma_Y$, the standard deviation, is the square root of the variance.

## A.2 Monte–Carlo integration

Monte–Carlo integration is a numerical integration technique based on random sampling. We independently draw $M$ input samples $\mathcal{X} = \{\boldsymbol{x}^{(1)}, \ldots, \boldsymbol{x}^{(M)}\}$ from the input distribution and compute the corresponding responses $\mathcal{Y} = \{y^{(1)}, \ldots, y^{(M)}\}$. Thus, $y^{(m)} = f(\boldsymbol{x}^{(m)})$ are realisations of the random variable $Y$, for $m = 1, \ldots, M$. The Monte–Carlo estimate of the expected value, given in Eq. (A.3), then is

$$\mathbb{E}\left[Y\right] = \frac{1}{M} \sum_{m=1}^{M} y^{(m)}. \tag{A.5}$$

Monte–Carlo estimates have a convergence rate of $\mathcal{O}(1/\sqrt{M})$ which is independent of the input dimension. Therefore, Monte–Carlo integration is especially suitable for high-dimensional problems.

## A.3 Numerical approximation of the conditional variances involved in the Sobol indices

The first-order Sobol index is defined as

$$S^i = \frac{\mathbb{V}_{X_i}\left[\mathbb{E}_{\mathbf{X}_{\sim i}}\left[Y|X_i\right]\right]}{\mathbb{V}\left[Y\right]}. \tag{A.6}$$

Hence, the question arises as to how the conditional variance $\mathbb{V}_{X_i}\left[\mathbb{E}_{\mathbf{X}_{\sim i}}\left[Y|X_i\right]\right]$ is estimated in practice. In the following, the notation $\int f(x_i)\,dx_i$ is used in place of $\int f(x_i)p(x_i)\,dx_i$ implying that the input parameters' probability density functions are embedded in the integrand $f(x_i)$ [1].

For the first-order index, we rewrite

$$\mathbb{V}_{X_i}\left[\mathbb{E}_{\mathbf{X}_{\sim i}}\left[Y|X_i\right]\right] = \mathbb{E}_{X_i}\left[\mathbb{E}^2_{\mathbf{X}_{\sim i}}\left[Y|X_i\right]\right] - \left(\mathbb{E}_{X_i}\left[\mathbb{E}_{\mathbf{X}_{\sim i}}\left[Y|X_i\right]\right]\right)^2 \tag{A.7}$$

$$= \int \mathbb{E}^2_{\mathbf{X}_{\sim i}}\left[Y|X_i\right]\,dx_i - \left(\int \mathbb{E}_{\mathbf{X}_{\sim i}}\left[Y|X_i\right]\,dx_i\right)^2. \tag{A.8}$$

The latter integral is $\left(\mathbb{E}\left[Y\right]\right)^2$, since

$$\int \mathbb{E}_{\mathbf{X}_{\sim i}}\left[Y|X_i\right]\,dx_i = \iint Y\,dx_i\,d\mathbf{x}_{\sim i} = \int Y\,d\mathbf{x}. \tag{A.9}$$

However, the former integral in Eq. (A.8) is computationally impractical. In a Monte–Carlo frame, such terms imply a double loop and thus a computational cost of $M^2$, with $M$ being the number of Monte–Carlo samples. Ishigami and Homma [2] rewrote this integral as

$$\int \mathbb{E}^2_{\mathbf{X}_{\sim i}}\left[Y|X_i\right]\,dx_i = \iiint f(X_1,\ldots,X_D)\cdot f(X'_1,\ldots,X_i,\ldots,X'_D)\,d\mathbf{x}_{\sim i}\,d\mathbf{x}'_{\sim i}\,dx_i \tag{A.10}$$

$$= \iint f(X_1,\ldots,X_D)\cdot f(X'_1,\ldots,X_i,\ldots,X'_D)\,d\mathbf{x}\,d\mathbf{x}'_{\sim i}, \tag{A.11}$$

where the last line is equal to the expected value of

$$F(X_1,\ldots,X_D,X'_1,\ldots,X_i,\ldots,X'_D) = f(X_1,\ldots,X_D)\cdot f(X'_1,\ldots,X_i,\ldots,X'_D), \tag{A.12}$$

and it hence can be computed using a single Monte–Carlo loop.

To estimate the first- and the total-order Sobol indices, following the best practices discussed by Saltelli et al. [1], we set up a triplet of matrices $\mathbf{A}$, $\mathbf{B}$ and $\mathbf{A}_\mathbf{B}^{(i)}$. First, we generate two independent sampling matrices $\mathbf{A}$ and $\mathbf{B}$ as

$$\mathbf{A} = \begin{pmatrix} x_{11} & x_{12} & \ldots & x_{1i} & \ldots & x_{1D} \\ x_{21} & x_{22} & \ldots & x_{2i} & \ldots & x_{2D} \\ \ldots \\ x_{M1} & x_{M2} & \ldots & x_{Mi} & \ldots & x_{MD} \end{pmatrix} \tag{A.13}$$

$$\mathbf{B} = \begin{pmatrix} x'_{11} & x'_{12} & \ldots & x'_{1i} & \ldots & x'_{1D} \\ x'_{21} & x'_{22} & \ldots & x'_{2i} & \ldots & x'_{2D} \\ \ldots \\ x'_{M1} & x'_{M2} & \ldots & x'_{Mi} & \ldots & x'_{MD} \end{pmatrix}. \tag{A.14}$$

To generate those sampling matrices $\mathbf{A}$ and $\mathbf{B}$, we use sequences of quasi-random numbers with low discrepancy, as for example the Sobol sequence [3]. We write $x_{ji}$ as a generic element: the index $j$ is the row index running from 1 to $M$, the number of Monte–Carlo samples, and similar the index $i$ is

the column index running from 1 to $D$, the number of input space dimensions. We now introduce the third matrix $A_B^{(i)}$ as

$$A_B^{(i)} = \begin{pmatrix} x_{11} & x_{12} & \cdots & x'_{1i} & \cdots & x_{1D} \\ x_{21} & x_{22} & \cdots & x'_{2i} & \cdots & x_{2D} \\ \cdots & & & & & \\ x_{M1} & x_{M2} & \cdots & x'_{Mi} & \cdots & x_{MD} \end{pmatrix}, \qquad (A.15)$$

where all columns are taken from matrix $A$ apart from the $i$-th column which is taken from matrix $B$. One sample $j$ in $B$ and the corresponding sample $j$ in $A_B^{(i)}$ have in common $x'_{ji}$ but differ in all other parameters $x_{\sim i}$.

Hence, we have to evaluate our model $f$ at all samples of the triplet $A$, $B$ and $A_B^{(i)}$: this results in $2M$ simulations needed for computing $f(A)$ and $f(B)$ plus $D \cdot M$ simulations needed for computing $f(A_B^{(i)})$. In sum, the computational cost is $M \cdot (D+2)$, with $M$ being sufficiently large (500 or higher).

## A.4 Algorithm

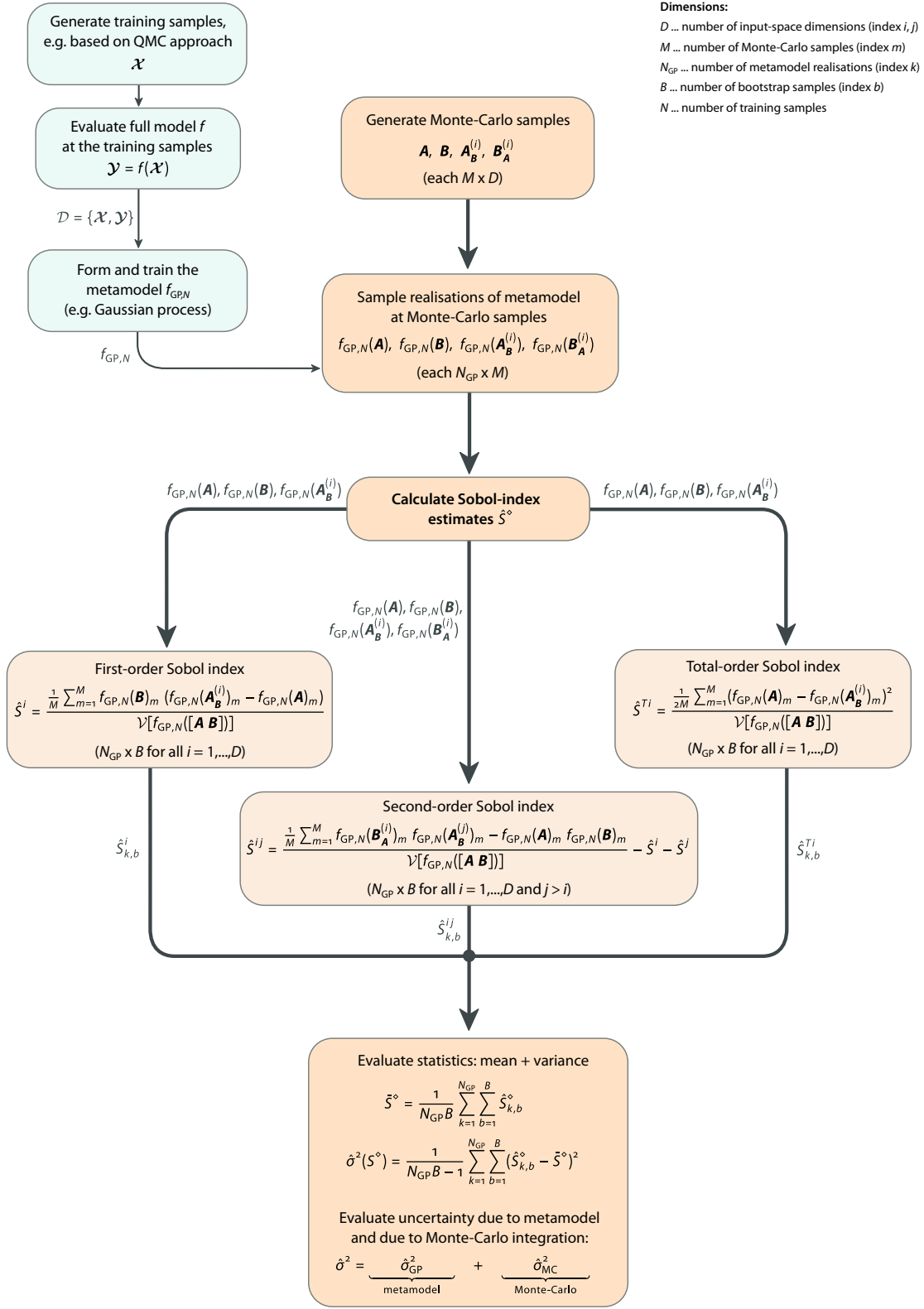

## A.5 Full results for the application to a model of arterial growth and remodelling

Fig. A.1 depicts the projection of the (trained) Gaussian process into the input-space dimensions including the projection of the mean $m_{\text{GP}i}(X_i)$, projected 95% confidence interval (CI), and training samples $\mathcal{D}$ for the diameter $d_{\max}$.

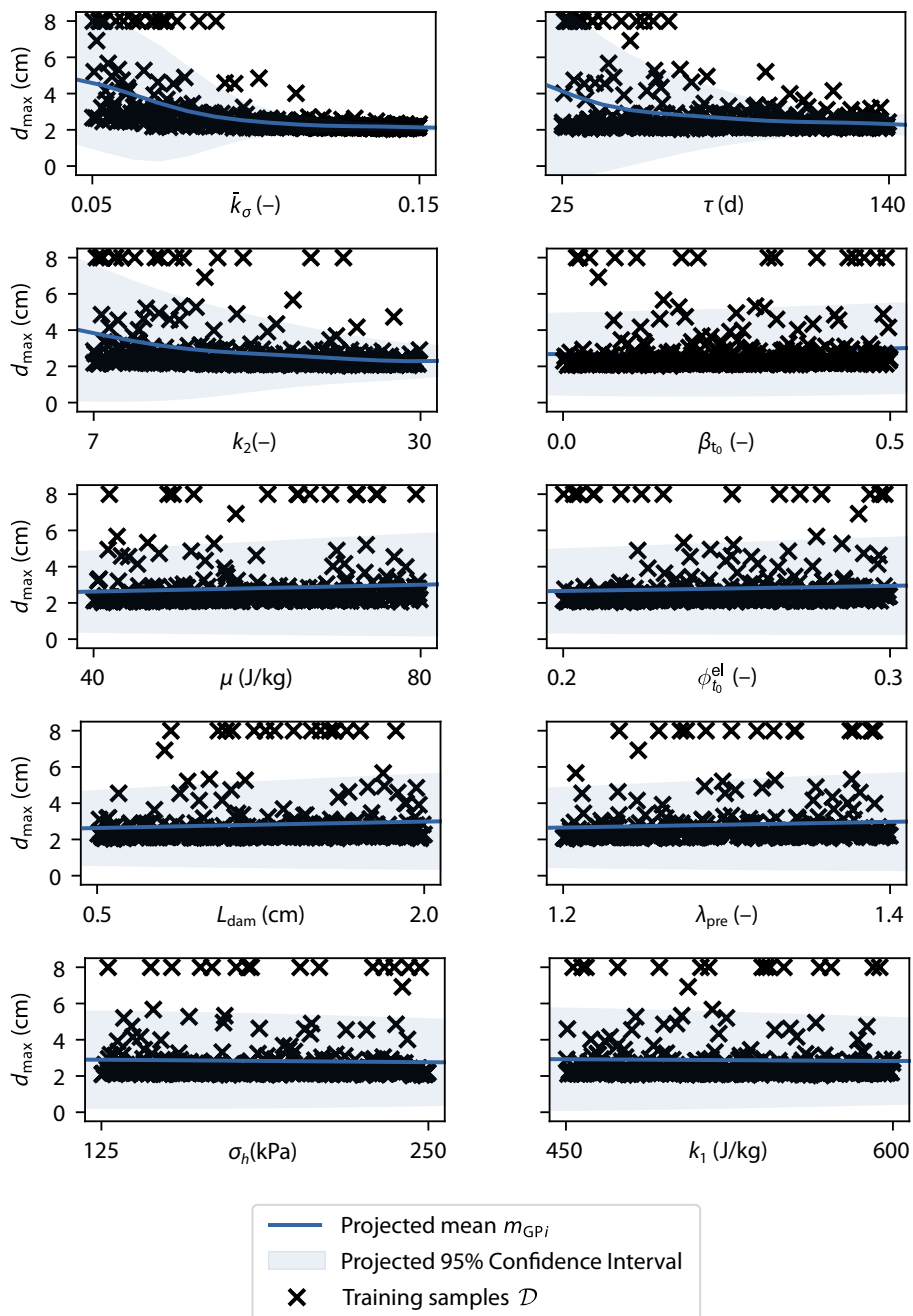

**Figure A.1: Gaussian process.** Projected mean $m_{\text{GP}i}(X_i)$, projected 95% confidence interval (CI) and training samples $\mathcal{D}$ for the diameter $d_{\max}$.

Fig. A.2 presents the estimates for the first and the total-order Sobol indices for increasing numbers of training samples. The results are based on $M = 10\,000$ Monte–Carlo samples, $N_{\mathrm{GP}} = 500$ realisations of the Gaussian process, and $B = 300$ bootstrap samples. We additionally include the 95% confidence intervals (due to the metamodel and the sum of CI due to the metamodel and Monte–Carlo integration).

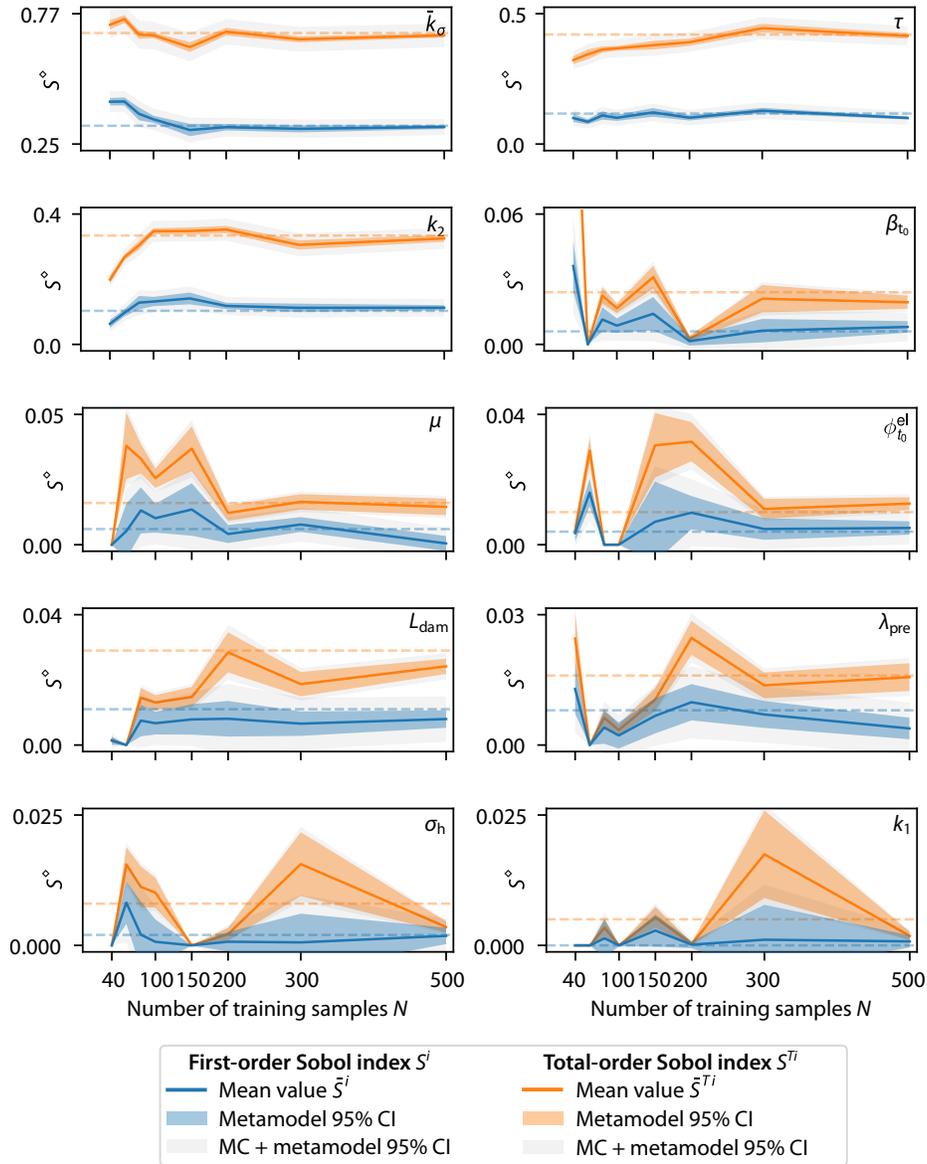

Figure A.2: **First-order and total-order Sobol indices and 95% confidence intervals (CI) for an increasing number of training samples**. We use $M = 10\,000$ Monte–Carlo samples, $N_{\mathrm{GP}} = 500$ realisations of the Gaussian process, and $B = 300$ bootstrap samples.

6

Further, we calculate the second-order Sobol indices $S^{ij}$, including both sources of uncertainty based on $M = 300\,000$ Monte–Carlo samples, $N_{\text{GP}} = 500$ realisations of the Gaussian process, $B = 300$ bootstrap samples, and $N = 300$ training samples. Fig. A.3 summarises the results together with the corresponding 95% confidence intervals. The remaining second-order Sobol indices are $S^{ij} < 0.0005$.

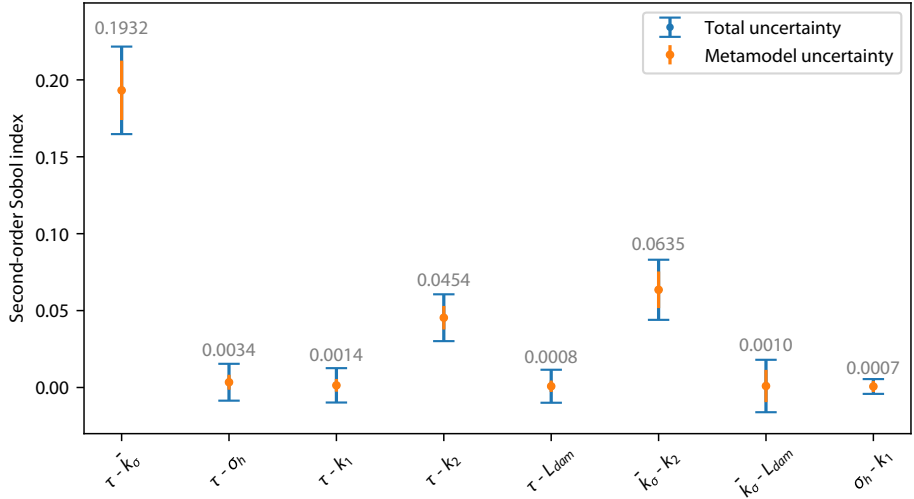

Figure A.3: **Second-order Sobol indices**. We include the 95% confidence intervals (due to the metamodel and the sum of CI due to the metamodel and Monte–Carlo integration).

Finally, we estimate the third-order Sobol indices for the parameter combination $\bar{k}_\sigma$, $\tau$ and $k_2$. The total-order Sobol index $S^{Ti}$ includes all contributions of parameter $X_i$

$$S^{Ti} = S^i + \sum_{j \neq i}^{D} S^{ij} + \cdots + S^{12\ldots D}. \tag{A.16}$$

Thus, summing up the first-order index $S^i$ plus all second-order indices $S^{ij}$ and comparing this sum to the total-order index $S^{Ti}$ reveals whether we expect third-order or even higher order interaction effects. For all three parameters $\bar{k}_\sigma$, $\tau$ and $k_2$, the total-order index is higher than the sum of first- and second-order indices

$$S^{Ti} > S^i + \sum_{j \neq i}^{D} S^{ij} \tag{A.17}$$

(see Table A.1). We therefore specifically calculate $S^{\bar{k}_\sigma \tau k_2}$, which describes the interaction of those three parameters. The third-order effect is estimated according the general form for the estimator as presented by [4] as

$$\hat{S}^{ijk} = \frac{\frac{1}{M}\sum \tilde{f}_{\text{GP},N}(\boldsymbol{B})_m \tilde{f}_{\text{GP},N}(\boldsymbol{A}_{\boldsymbol{B}}^{(ijk)})_m - \frac{1}{M}\sum \tilde{f}_{\text{GP},N}(\boldsymbol{B})_m \frac{1}{M}\sum \tilde{f}_{\text{GP},N}(\boldsymbol{A}_{\boldsymbol{B}}^{(ijk)})_m}{\frac{1}{M}\sum \tilde{f}_{\text{GP},N}(\boldsymbol{B})_m^2 - \left(\frac{1}{M}\sum \tilde{f}_{\text{GP},N}(\boldsymbol{B})_m\right)^2}$$
$$- \hat{S}^{ij} - \hat{S}^{ik} - \hat{S}^{jk} - \hat{S}^i - \hat{S}^j - \hat{S}^k, \tag{A.18}$$

where all sums are $\sum_{m=1}^{M}$. $\boldsymbol{A}_{\boldsymbol{B}}^{(ijk)}$ is build similar to $\boldsymbol{A}_{\boldsymbol{B}}^{(i)}$, i.e. one sample $(\boldsymbol{B})_m$ and the corresponding sample $(\boldsymbol{A}_{\boldsymbol{B}}^{(ijk)})_m$ have in common $x_i, x_j$ and $x_k$ but differ in all other parameters $\boldsymbol{x}_{\sim ijk}$ (resample all but $x_i, x_j$ and $x_k$ for all $M$ Monte–Carlo samples). Using $M = 500\,000$ Monte–Carlo samples, $N_{\text{GP}} = 500$ realisations of the Gaussian process, $B = 300$ bootstrap samples, and $N = 500$ training samples results in $S^{\bar{k}_\sigma \tau k_2} = 0.06$ and 95% confidence intervals of $\text{CI}_{\text{MC}} = 0.011$ and $\text{CI}_{\text{MC}} = 0.009$. Adding this single

third-order index to the previous sum of first and second-order indices (see Table A.1), the result is (almost) equal to the corresponding total-order index. We thus identified (almost) all contributions of first and higher order.

Table A.1: **Sum of Sobol indices.** Values are rounded to two decimal places. $\sum = S^i + \sum S^{ij} + S^{\bar{k}_\sigma \tau k_2}$.

| Parameter | $S^i$ | $\sum S^{ij}$ | $S^{\bar{k}_\sigma \tau k_2}$ | $\sum$ | $S^{Ti}$ |
|---|---|---|---|---|---|
| $\bar{k}_\sigma$ | 0.33 | 0.25 | 0.06 | 0.64 | 0.67 |
| $\tau$ | 0.14 | 0.25 | 0.06 | 0.45 | 0.46 |
| $k_2$ | 0.10 | 0.11 | 0.06 | 0.27 | 0.27 |



\end{document}